\documentclass[twocolumn]{aastex631}

\usepackage{xcolor}
\usepackage{multirow}
\usepackage{hyperref}
\usepackage{amsmath}
\hypersetup{linkcolor=red,citecolor=blue,filecolor=cyan,urlcolor=blue}
\shorttitle{Dwarf AGN in IllustrisTNG}
\shortauthors{M. T. Kristensen et al.}
\graphicspath{{./}{figures/}}

\begin{document}

\title{Merger Histories and Environments of Dwarf AGN in IllustrisTNG}

\correspondingauthor{Mikkel Theiss Kristensen}
\email{m.t.kristensen-2018@hull.ac.uk}

\author[0000-0002-4802-0452]{Mikkel Theiss Kristensen}
\affiliation{E.A Milne Centre, University of Hull,
Cottingham Road,
Hull, HU6 7RX, UK}

\author[0000-0002-3963-3919]{Kevin Pimbblet}
\affiliation{E.A Milne Centre, University of Hull,
Cottingham Road,
Hull, HU6 7RX, UK}


\author[0000-0003-4446-3130]{Brad Gibson}
\affiliation{E.A Milne Centre, University of Hull,
Cottingham Road,
Hull, HU6 7RX, UK}

\author[0000-0001-5703-7531]{Samantha Penny}
\affiliation{Institute of Cosmology and Gravitation, University of Portsmouth,
Dennis Sciama Building, Burnaby Road,
Portsmouth, PO1 3FX, UK }

\author{Sophie Koudmani}
\affiliation{Institute of Astronomy, University of Cambridge,
Madingley Road,
Cambridge, CB3 0HA, UK}
\affiliation{Kavli Institute for Cosmology, University of Cambridge,
Madingley Road,
Cambridge, CB3 0HA, UK}

 
  \begin{abstract}
   The relationship between active galactic nuclei activity and environment has been long discussed, but it is unclear if these relations extend into the dwarf galaxy mass regime -- in part due to the limits in both observations and simulations.
   We aim to investigate if the merger histories and environments are significantly different between AGN and non-AGN dwarf galaxies in cosmological simulations, which may be indicative of the importance of these for AGN activity in dwarf galaxies, and whether these results are in line with observations.
   Using the IllustrisTNG flagship TNG100-1 run, 6\,771 dwarf galaxies are found with 3\,863 ($\sim$57 per cent) having some level of AGN activity. In order to quantify `environment', two measures are used: 1) the distance to a galaxy's 10th nearest neighbour at 6 redshifts and 2) the time since last merger for three different minimum merger mass ratios. A similar analysis is run on TNG50-1 and Illustris-1 to test for the robustness of the findings.
   Both measures yield significantly different distributions between AGN and non-AGN galaxies; more non-AGN than AGN galaxies have long term residence in dense environments while recent ($\leq 4 \text{ Gyr}$) minor mergers are more common for intermediate AGN activity. While no statements are made about the micro- or macrophysics from these results, it is nevertheless indicative of a non-neglible role of mergers and environments.
   \end{abstract}


%

\section{Introduction}
An important part of galaxy evolution is the co-evolution of the central black hole and the central bulge. The black hole mass and the luminosity and mass of the bulge follow a tight correlation for classical bulges and elliptical galaxies \citep{Marconi2003, Gueltekin2009, Alexander2012, Volonteri2012, Kormendy2013}. Although the evolution of the two components are closely linked, divergences from the trend suggest that the bulge and supermassive black hole do not follow the exact same channels of evolution \citep[e.g][]{Simmons2017}. 

Conventionally, the growth and even formation of elliptical galaxies and classical bulges are believed to be mainly through mergers \citep{Kormendy2013}. Galaxies in the process of merging often show undermassive super massive black holes (SMBHs) and several studies do not find a significant link between mergers and the so-called active galactic nucleus phase (AGN; a phase where the SMBH grows through gas accretion) both observationally \citep[][]{Villforth2016, Simmons2017, Sugata2019, Smethurst2019} and in simulations \citep{Steinborn2018, Martin2018, Ricarte2019} lending credit to the fact that the two components evolve somewhat independently -- although some studies find merger activity and type of merger linked to the type and strength of AGN \citep[e.g][]{Satyapal2014, Simmons2017, Donley2018, Shah2020}.

The abovementioned growth phase, the AGN phase, is when a SMBH is accreting gas and material. In this phase, matter is being driven to the central region of the galaxy and is being deposited onto the black hole. There are a number of mechanisms thought to be able to drive gas and dust to the center, either through internal processes such supernova feedback or dynamic friction (commonly referred to as secular evolution) or through external ones such as ram pressure stripping \citep{Gunn1972}, galaxy interactions \citep{Moore1996} or accretion from the intergalactic medium (i.e environmental effects). 

However, studies examining black hole growth have primarily been focused on intermediate mass galaxies or high mass galaxies \citep[understood as $M_{*} \geq 10^{10} M_{\odot}$, see e.g][]{DiMatteo2005, Bower2006, Sijacki2009, Amiri2019}. While these galaxies are easier to study with their greater brightness and size, they are not necessarily representative of lower mass galaxies and their evolution, which is evidenced by the difference in susceptibility to different quenching mechanisms between mass regimes  of galaxies \citep{Peng2010, Peng2012, Geha2012}. 

In order to decrypt the potential differences between populations, a good statistical basis is required. Obtaining a large number of sources requires large scale surveys. However, observations have traditionally had to choose between either large field of view or faint magnitude limits where deep surveys do not yield a high number of sources, but it would include the faint ones (i.e the low mass galaxies) while wide surveys would include many sources but little to no low mass galaxies. As prominent examples of this is the Sloan Digital Sky Survey \citep[SDSS][]{York2000} DR16 which covers 14\,555 square degrees to a limiting magnitude of around 22 (\textit{ugriz} bands) while the UltraVISTA \citep[][]{McCracken2012} covers 1.5 square degrees to a limiting magnitude of around 25 (\textit{Y} band).
 
Similarly, in large-scale cosmological simulations, resolving dwarf galaxies requires low particle masses and/or low mass cells which results in a high particle/cell count, which is computationally expensive and thus not feasible to pursue. As an example, one of the largest and earliest cosmological scale simulations, the Millenium Simulation \citep{Springel2005Mil}, used $2\,160^3$ dark matter particles with an individual particle mass of $8.6 \times 10^8 h^{-1}$ M$_{\odot}$ -- roughly half the dark matter mass of the Small Magellanic Cloud \citep[][]{DiTeodoro2019}. However, progress being made with high resolution boxes such as TNG50 \citep[][]{Pillepich2019, Nelson2019a} and zoom-in simulations of large regions such as NewHorizon \citep[][]{Dubois2020}.

Evidently, the low mass regime of galaxies is a relatively under explored field, and even more so when considering dwarf galaxies with AGN characteristics. However, with advances in both observations and simulations, this undertaking has become more feasable. Observationally, \citet{Greene2004, Greene2007, Reines2013, Sartori2015, Mezcua2016} were some of the first ones to look at dwarf AGNs and central black holes on a large scale. \citet{Baldassare2017} examined the x-ray and UV properties of AGNs in nearby dwarf galaxies, again expanding dwarf AGNs into a new realm. Detailed studies on the impact of AGN in dwarf galaxies is also feasible now -- from outflows \citep{ManzanoKing2019, Liu2020} to feedback and gas kinematics \citep{Dashyan2018, Penny2018, Kaviraj2019, Reines2020}

Cosmological simulations, too, now reach somewhat resolved dwarf galaxies ($M_{*} \leq 5 \times 10^{9} M_{\odot}$) such as IllustrisTNG with baryonic particle masses between $5.7 \times 10^4 - 7.6 \times 10^6 \text{ M}_{\odot}/h$ (in TNG50-1 and TNG300-1, respectively). Although smaller scale simulations (local group size) with dwarf galaxies have been around for longer \citep[e.g][]{Wadepuhl2011} and are still being refined today \citep{Trebitsh2018,Barai2019, Koudmani2019, Bellovary2019, Sharma2020}, only now is the emphasis on the effect of AGNs and black hole growth.

One of the keystone subjects of AGNs is under what circumstances they are found and what triggers their activity. Examples of such questions are whether field galaxies more frequently host AGNs, whether mergers and tidal interactions are the main culprit of triggering AGN activity, or what the effect of a dense environment is. While a connection between density of a galaxy's environment and its star formation rate has been established \citep{Baldry2006, Peng2010, Davidzon2016, Penny2016}, the environment-AGN connection is disputed \citep{Yang2018, Smethurst2019, Kristensen2020}. 

However, a connection between strong AGN and merger activity has been found \citep[e.g][]{Steinborn2018, Ellison2019, Kaviraj2019, Marian2020} suggesting that external factors are not without a say. The lack of an apparent connection to environment may be due to a time delay between the conditions that triggered AGN activity and when the AGN activity turned on \citep[e.g][]{Hopkins2012, Pimbblet2013, Kristensen2020}

Observationally, past environments and events are hard -- if not impossible -- to find unless morphological disturbances are still present. Some promise has been found using integrated field unit (IFU) spectroscopy where \citet{Penny2018} found kinematically offset cores in a sample dwarf AGN galaxies. However, simulations retain the complete environmental history and past mergers -- tracers of which are erased over time in real galaxies. 

This study aims to test whether or not the current and past environments of a sample low-z dwarf AGN galaxies are different from those of a matched control sample with no AGN activity. The environment is examined in the IllustrisTNG simulation (more specifically, the TNG100-1 run), and observational data from the NASA-Sloan Atlas (NSA) is also included for comparative purposes. The AGN samples and the control samples are compared against each other using a Monte Carlo Kolmogorov-Smirnov (KS) test suite following a similar procedure as \citet{Kristensen2020}.

This paper is organised as follow: Section \ref{sec:dataandmethods} describes the data used, the sample selection criteria, and the environmental measures used. Section \ref{sec:results} contains the different distributions of the parameters of the the different samples along with the results from the KS-testing. Caveats and discussion of the results follow in Section \ref{sec:discussion} and the findings are summarised in Section \ref{sec:summary}.  This study assumes the same cosmology as IllustrisTNG, namely a $\Lambda$-CDM Universe
with $\Omega_{\Lambda,0} = 0.6911$, $\Omega_{m,0} = 0.3089$, $\Omega_{b,0} = 0.0486$, and $h=0.6774$

\section{Data and Methods}
\label{sec:dataandmethods}
This section will describe the data used and details about the analysis carried out on said data. The data used is mostly simulation data from the IllustrisTNG project using their 75 Mpc$/h \sim 106.5$ Mpc simulation (TNG100-1) with some observational data from the NASA-Sloan Atlas (NSA, details of this data set can be found in \citet{Kristensen2020}, but can be summarised as SDSS dwarf galaxies with M$_{*} \leq 3\times 10^9$ M$_{\odot}$, $z \leq 0.055$) included for comparison purposes. The samples are first found using dwarf galaxy selection criteria similar to \citet{Kristensen2020} combined with simulation specific requirements, and that sample is then subdivided into subsamples according to AGN selection criteria based on Eddington ratios. Finally, a number of environmental measures are found for all dwarf galaxies, and those properties are then compared between the different subsamples using a Monte Carlo Kolmogorov-Smirnov testing suite.

\subsection{IllustrisTNG and Illustris}
\label{sec:data:sims}
The Illustris `The Next Generation' (IllustrisTNG) simulation is the successor to the original Illustris simulation \citep{Vogelsberger2014, Genel2014, Sijacki2015} with updated and new physics and refinements over the original. The simulations are evolved with the AREPO code \citep{Springel2010}, and consists of three different runs (TNG50, TNG100, and TNG300), though only TNG100 is used for this analysis. The number indicates the physical box size, and for TNG100, side lengths of the box are 75 Mpc$/h \sim 106.5$ Mpc, with $h=0.6774$. More specifically, the TNG100-1 run is used which has $1\,820^3$ dark matter particles with a mass of $7.5 \times 10^6 M_{\odot}$ and a baryonic mass of $1.4 \times 10^6 M_{\odot}$.

Of particular interest is the evolution and modelling of supermassive black holes. As described in \citet{Weinberger2018}, friends of friends (FoF) groups are identified on the fly on dark matter particles and a SMBH of mass $1.18 \times 10^6$ M$_{\odot}$ is seeded whenever a FoF halo exceeds a total mass threshold of $7.38 \times 10^{10}$ M$_{\odot}$ and does not yet contain a SMBH. This does mean that some low mass subhalos in very dense environments are not seeded a black hole. The subhalos without black holes are not included in the analysis. A thorough discussion of this bias can be found in Section \ref{sec:BHDiscussion}.

The mass accretion of the black holes are reliant on the local environment -- more specifically, it is a Bondi-based accretion prescription, which relies on the ambient sound speed and ambient density. It is not boosted as in other simulations \citep[e.g ][]{Springel2005}, which gives more validity to the environmental analysis since the accretion rate is based only on the physical space and processes surrounding the black hole. There are, however, caveats with the BH modelling. Therefore, to check the validity of the results obtained from TNG100-1, the TNG50-1 and Illustris-1 runs are used for comparison purposes.

\begin{table}
\caption{Overview of relevant simulation parameters}             

\label{tab:sim_compare}      
\centering                          
\begin{tabular}{l l l l l}        
\hline\hline                 
Simulation & L$_{\text{box}}$ [ckpc/h] & $m_{\text{DM}}$ & $m_{\text{gas}}$ & $m_{\text{BH seed}}$ \\    
\hline                        
  TNG100-1      & 75000 & 750   & 140   & 80/h \\      
  TNG50-1       & 35000 & 45    & 8.5   & 80/h \\
  Illustris-1   & 75000 & 630   & 130   & 10/h \\
\hline                                   
\end{tabular}
\tablecomments{First column is the simulation name, second one is the corresponding side length given in units of comoving kpc/h. Third column is the mass of a dark matter particle followed by gas cell/particle mass and lastly is the mass of the seeded black hole particle. Masses are in $10^4 \text{ M}_{\odot}$ for easy comparison.}
\end{table}

The main differences in resolution and particle masses between the three simulations can be found in Table~\ref{tab:sim_compare}, but to summarise: TNG100-1 and Illustris-1 have roughly the same box size and particle masses, though BH seeds are lighter in Illustris but BH accretion is boosted. BH feedback mechanisms are also slightly different and a more detailed discussion can be found in \citet{Pillepich2018} and references therein. TNG100-1 and TNG50-1 differ only in box size and particle masses -- their BH prescriptions are the same.

\begin{figure}
	\centering
	\includegraphics[width=\linewidth]{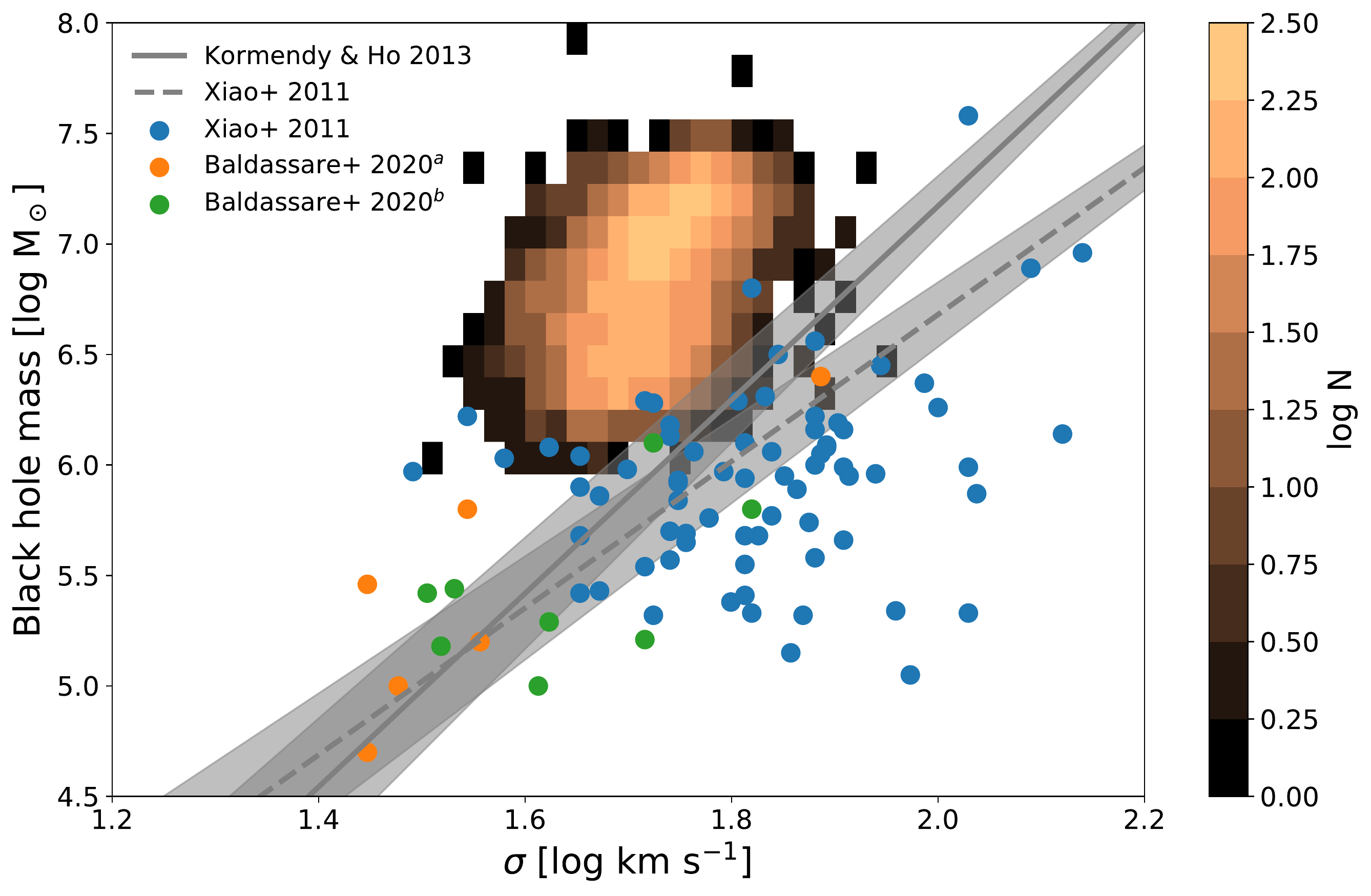}
	\caption{Black hole mass versus stellar velocity dispersion, $\sigma$. Two relations are plotted (\citet{Xiao2011} dashed line, \citet{Kormendy2013} solid line) with their intrinsic scatter. Observations of black holes in dwarf galaxies (M$_{*}$ between 8.5-9.5 log M$_{\odot}$) from \citet{Xiao2011} are in blue and \citet{Baldassare2020} are in orange and green -- \textit{a}: from previous work, \textit{b}: from \citet{Baldassare2020} study. The copper 2D histogram shows TNG100-1 galaxies }
	\label{fig:bh-sigma_rel}
\end{figure}

Both TNG and Illustris have caveats and limitations in the BH modelling \citep{Weinberger2018, Li2020,Terrazas2020,Habouzit2021}. \citet{Habouzit2021} remarks that current cosmological simulations -- Illustris(TNG) and Horizon-AGN, EAGLE, and SIMBA -- struggle to produce the diversity of BHs observed in the local Universe. A strong caveat of TNG BH modelling is the seeding mass of $\sim 10^6 \text{M}_{\odot}$, which is $\sim 0.5$ dex more massive than currently observed BH in dwarf galaxies \citep[e.g][]{Xiao2011, Kormendy2013, Reines2013,Moran2014, Reines2015,ManzanoKing2020,Baldassare2020}. This effect can be seen in Figure \ref{fig:bh-sigma_rel}, which shows the black hole-velocity dispersion relationship of both TNG100-1 subhalos and observed black holes in dwarf galaxies in \citet{Xiao2011} and \citet{Baldassare2020}. 
However, our AGN selection relies on the Eddington ratios (see Section \ref{sec:AGNSelection}, which scales with $M_{\text{BH}}$ and not accretion rates, which scales with $M_{\text{BH}}^2$, lessening the importance of the bias of overmassive SMBHs. Therefore, the results should also be seen as a comment on the simulation physics. Nevertheless, the impact of seed mass is checked by including Illustris-1, which has a lower seed mass than the TNG runs.

Merger trees and thus merger data are from the SubLink algorithm \citep{RodriguezGomez2015}.
Colours are from \citet{Nelson2018} which are calculated from the stellar particles in a subhalo by summing the luminosities of the particles and applying a dust attenuation model. The response function is modelled to SDSS photometry. For Illustris-1 and TNG50-1, dust corrected measurements are not available and thus just the sum of luminosities of the stellar particles are used.

\subsection{Dwarf galaxy selection}
\label{sec:dwarfSelection}
\begin{figure*}
	\centering
	\includegraphics[width=0.95\linewidth]{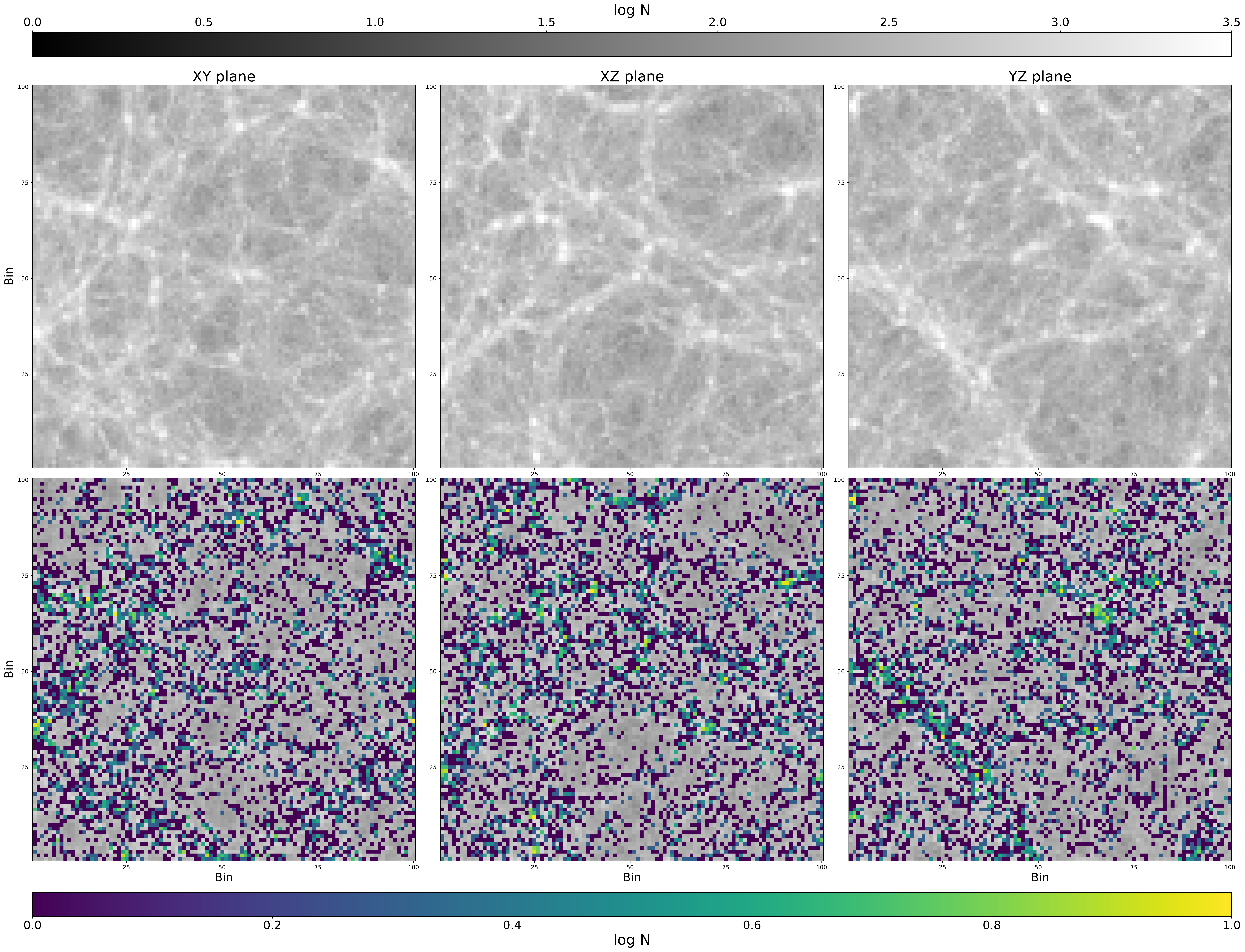}
	\caption{Spatial distribution of all subhalos only in top row and with selected dwarf galaxies bottom row projected onto three different planes (XY, XZ, and YZ plane). The gray scale background number density plot includes all subhalos while the coloured distribution is for dwarf galaxies. The data is split into 100 bins on each axis resulting in a bin size of $0.75 \times 0.75$ Mpc/\textit{h}}
	\label{fig:density_dwarfs}
\end{figure*}
During the simulation run, group catalogues are computed using friends-of-friends (FoF) and Subfind algorithms at each snapshot on dark matter particles. For this study, dwarf galaxies are selected from the $z=0$ (snapshot 99) group catalogue, and a number of requirements are imposed to ensure the selected dwarf galaxies are of acceptable quality. Working with dwarf galaxies means working near the resolution limit of the simulation, and extra care has to be taken. 

Simulation specific requirements are that subhalos are required to have a dark matter component and a black hole. An upper stellar mass limit $M_{*} = 3 \times 10^{9}$ M$_{\odot}$ is also imposed as this follows observational definitions. To resemble observations further, the stellar mass used is the mass enclosed in 2 times the stellar half mass radius, \texttt{PartType4} of \texttt{SubhaloMassInRadType}. Similarly, a lower stellar mass limit of $10^9$ M$_{\odot}$ is also used to resemble large scale observational surveys where this the effective mass limit \citep[see e.g][ Figure 11 for a mass distribution of local SDSS dwarf galaxies]{Kristensen2020}  and to ensure well resolved galaxies. This is in line with other studies on dwarf galaxies in simulations \citep{Sharma2020, Fattahi2020, Dickey2021, Koudmani2021, Reddish2021, Jahn2021}. 

These requirements can be summarised as follows:
\begin{enumerate}
    \item $1 \times 10^{9} \leq$ $M_{*,\text{2HM}} \leq 3 \times 10^{9}$ M$_{\odot}$
    \item M$_{\text{DM}} > 0$ M$_{\odot}$
    \item M$_{\text{BH}} > 0$ M$_{\odot}$
\end{enumerate} 

A projected number density of dwarf galaxies on three planes can be seen in Figure \ref{fig:density_dwarfs}. These requirements yield 6\,771 dwarf galaxies.

The reasoning behind requiring a black hole is due to large number of dwarf galaxies with no black holes ($N=1\,001$, assuming same requirements for regular dwarf galaxies except M$_{\text{BH}} = 0$ M$_{\odot}$). This requirement is also used in other studies \citep[e.g][]{Koudmani2021}. For these galaxies' real life counterparts, there is no reason to assume they should not have a black hole as seeding mechanisms are still very unconstrained. It comes down to the BH seeding mechanism in IllustrisTNG, which leaves the simuated galaxies without a BH. Thus, they are unable to host AGN activity and will skew the AGN to non-AGN distributions that will be described in Section \ref{sec:AGNSelection}. A discussion of this effect can be found in Section \ref{sec:BHDiscussion}. 

\subsubsection{Divergence from observational dwarf galaxies}
\begin{figure}
	\centering
	\includegraphics[width=0.95\linewidth]{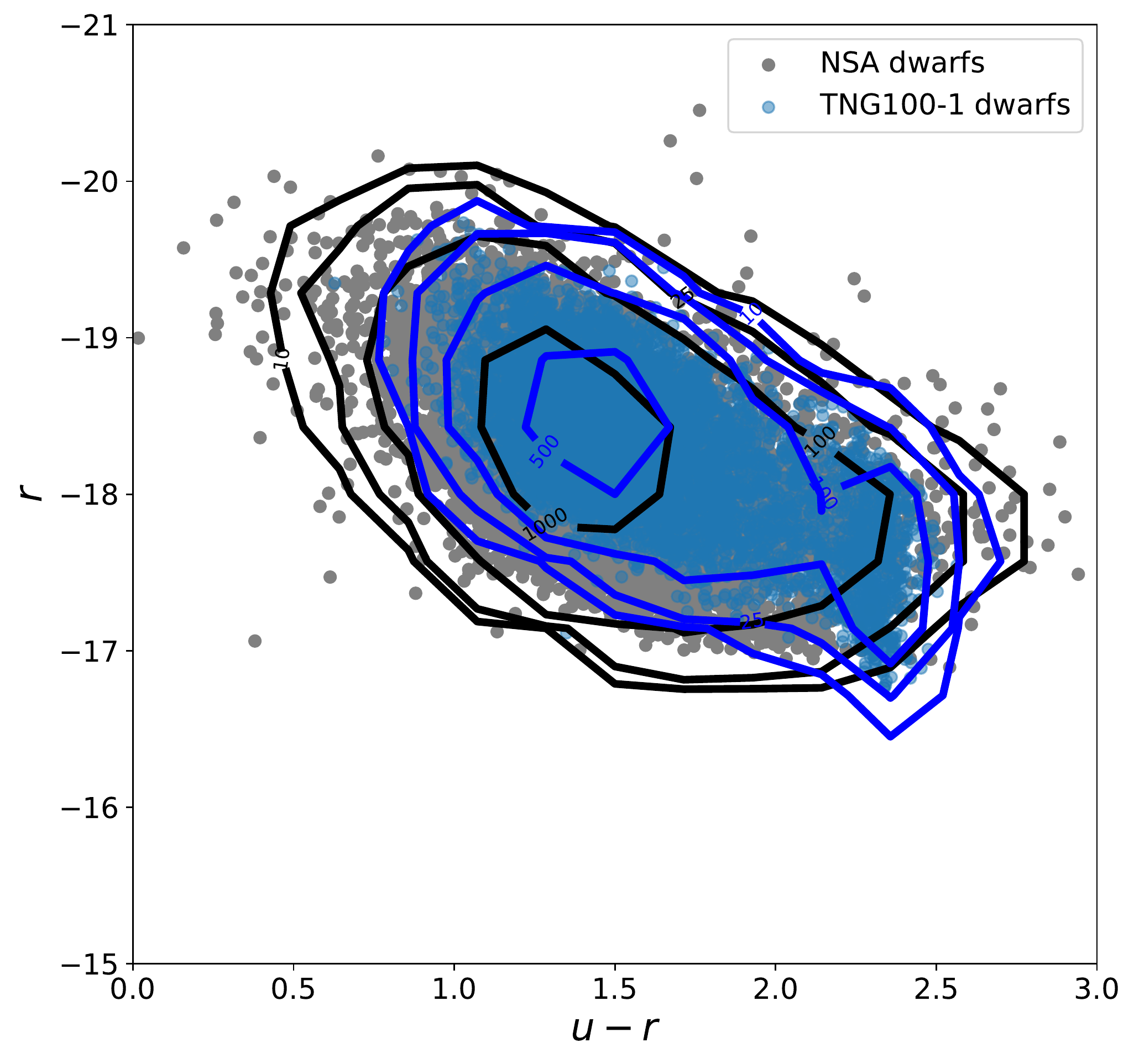}
	\caption{Colour-magnitude ($u-r$ colour vs $r$ magnitude) diagram showing SDSS dwarfs compared to TNG100 dwarfs with same mass selection criteria. Grey dots and black contour lines are NSA data while blue dots and contour lines are on dwarf subhalos from TNG100. The contour levels are at different levels between the samples since the sample sizes are different.}
	\label{fig:colour-mag_obs-sim}
\end{figure}

A concern when using observational data is bias towards low surface brightness galaxies not picked up due to observational constraints. This concern is not relevant when selecting subhalos from simulations since even the lowest surface brightness galaxies will be selected. Figure \ref{fig:colour-mag_obs-sim} shows a colour-magnitude diagram of TNG100-1 and NSA dwarfs. While the majority of dwarf galaxies and subhalos are centered around roughly the same values ($u-r = 1.3$, $r=-18.5$), NSA sources are more spread out -- especially towards bluer galaxies. Consequently, some bias exists in observational data but we do not attempt to correct for it since there are also uncertainties of the colour modelling of TNG100. The details of the colour model and subsequent bias of TNG are described in \citet{Nelson2018}.

\subsection{AGN selection} 
\label{sec:AGNSelection}

\begin{table*}
    \centering
    
    \caption{Number of subhalos for each AGN selection criteria.}
    
    \label{tab:agnnumbers}
    \begin{tabular}{lccccc}
        \hline\hline
        & Dwarfs & NOT & Weak & Intermediate & Strong \\
        Simulation & Eddington ratio & $\lambda < 0.005$ & $0.005 \leq \lambda < 0.01$ & $0.01 \leq \lambda < 0.1$ & $ \lambda \geq 0.1$\\ 
        
        \hline
        TNG100-1 & 6\,771 (100 \%) & 2\,908 (42.95 \%) & 988 (14.59 \%) & 2\,821 (41.66 \%) & 54 (0.80 \%) \\
        TNG50-1 & 1\,003 (100 \%) & 417 (41.58 \%) & 247 (24.63 \%) & 337 (33.60 \%) & 2 (0.20 \%) \\
        Illustris-1 & 10\,914 (100 \%) & 10\,402 (95.31 \%) & 226 (2.07 \%) & 254 (2.33 \%) & 32 (0.30 \%) \\
        \hline
    \end{tabular}
    \tablecomments{The total number is \textit{N} and how large a percentage of the total dwarf galaxy population is given in percentage in parenthesis. $\lambda$ is the Eddington ratio.}
\end{table*}

AGNs are selected from their Eddington ratios. The Eddington ratio is given as:

\begin{equation}
\begin{split}
    \lambda &= \dot{M}/\dot{M}_{\text{Edd}}, \\
    \dot{M}_{\text{Edd}} &= \frac{4\pi GM_{\text{BH}}m_p}{\varepsilon_{r} \sigma_{T} c} , \\
    \dot{M} &= \alpha 4\pi G^{2} M_{\text{BH}}^{2} \rho /c_s^3
\end{split}
\end{equation}

where $\dot{M}$ is the black hole mass accretion (Eddington limited Bondi accretion), $G$ is the gravitational constant, $M_{BH}$ is the mass of the black hole, $m_{p}$ is the proton mass, $\varepsilon_{r} = 0.2$ is the black hole radiative effeciency, $\sigma_{T}$ is the Thompson cross section, $c$ is the speed of light, $\alpha = 1$, $\rho$ is the local comoving gas density, and $c_s$ is the speed of sound in the local gas cells.The black hole mass and its accretion rate are based on the prescription in \citet{Springel2005} and are both available from the group catalogues and described in detail in Section 2.3 of \citet{Weinberger2018}.

Three Eddington ratios are used for AGN selection splits: $\lambda = 0.005, 0.01, 0.1$.  Table~\ref{tab:agnnumbers} contains the size of the samples along with their classification names; weak, intermediate, and strong. This follows similar selection as e.g \citet{Bhowmick2020, McAlpine2020}. Since the $\lambda \geq 0.1$ selection yields so few objects, it is difficult to draw convincing statistical conclusions for this sample. However, they are still kept for some analysis parts but are excluded in some plots and conclusions. Non-AGN (also referred to as the 'NOT' sample) are defined as dwarf galaxies with $ \lambda < 0.005$ and consists of 2\,908 (42.95 per cent) sources. An overview can be found in Table~\ref{tab:agnnumbers} which also contains an overview for TNG50-1 and Illustris-1. The discrepency in AGN fraction between TNG and Illustris is ascribed to the difference in BH-modelling and overmassive SMBH in TNG, which yields overly efficient black hole accretion (see Section \ref{sec:data:sims}).

The nature of Eddington ratio selected AGN are expected to be different than observationally chosen ones such as those in \citet{Kristensen2020}, which relies on optical emission line ratios. In fact, even AGN selected by different observational methods are expected to be different in nature from each other \citep[e.g][]{Ji2021}. AGNs selected by optical emission lines tend to not pick up obscured or low luminosity AGN -- attributes that do not matter when using an Eddington ratio selection. Following the classification scheme above on an observational study by \citet{Baldassare2017}, out of 12 dwarf AGN galaxies, 2 would be low intensity, 4 would be intermediate intensity, and 4 would be considered high intensity AGN. As such, observations appear to be biased towards high intensity/luminosity AGN.

Nevertheless, while simulations may include more low intensity AGN compared to observations, this bias is not important for the most part since comparisons between subsamples are within the same simulation runs, which means that compared galaxies have the overmassive black holes and boost factors. However, it does mean that direct comparisons to observations and between simulations need to be done cautiously.

\subsubsection{BH comparisons between simulations}
\label{sec:data:bh_comparison}
\begin{figure*}
	\centering
	\includegraphics[width=0.95\linewidth]{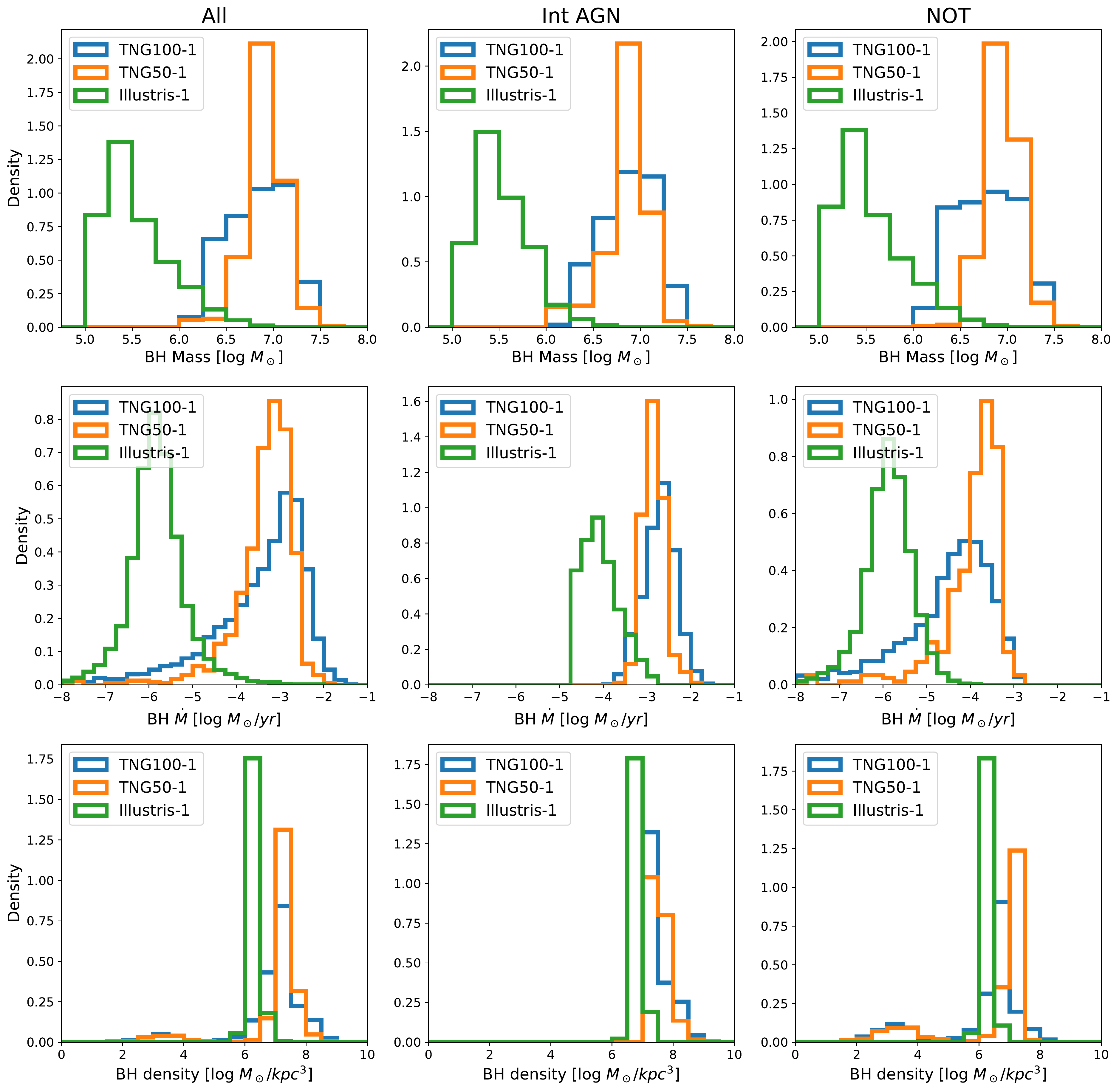}
	\caption{BH comparison histograms between simulations. Top row shows BH mass, middle row displays BH accretion rate, and the bottom row shows the density of local comoving gas of the BH. The columns display the full sample in the first row, Int AGN in the second column, and lastly non-AGN in the third column. TNG100-1 BH are in blue, TNG50-1 are orange, and Illustris-1 is green.}
	\label{fig:BHComparison}
\end{figure*}

Since TNG100 and TNG50 have the same BH modelling while Illustris runs follow slightly different prescriptions \citep[see][ Table 1 for a complete overview of the differences]{Pillepich2018}, some differences are expected between the AGN populations of the different simulations. Table \ref{tab:agnnumbers} shows a much lower AGN fraction for Illustris simulations ($\sim 4.7$ per cent) compared to TNG runs ($\sim 58$ per cent).

There are also differences in BH properties such as BH mass, accretion rate, and density of local comoving gas. Figure \ref{fig:BHComparison} shows that, as expected, the Illustris-1 BH mass is lower (by 1.5-2.0 dex). The BH accretion rate is also lower -- in fact by even more orders of magnitude than the black hole mass despite being boosted by a factor of $\alpha = 100$. This is not too surprising since the Bondi accretion rate is proportional to M$_{\text{BH}}^{2}$. Nevertheless, this is indicative of that AGN populations between simulations are different; TNG has more dwarf subhalos appearing as AGN because of a more efficient accretion, while subhalos sharing similar characteristics in Illustris may not have reached the $\lambda \geq 0.005$ threshold to be considered AGN. Further differentiating the accretion rates is the BH density -- the gas density from which accretion is calculated. In Illustris, it is only the parent gas cell while TNG calculates it from nearby gas cells (evaluated over a sphere enclosing certain number of neighbours \citep[where the neighbout number is scaled with the mass resolution of the simulation, see][ for details]{Weinberger2018, Pillepich2018}). Furthermore, TNG simulations have magnetohydrodynamic modelling unlike Illustris \citep{Pillepich2018}, which further changes the properties of the accretion gas cells.

Between TNG100 and TNG50, the differences are subtle. They roughly have the same average BH mass but with TNG50 being more concentrated (TNG100: $8.70 \pm 6.10 \times 10^{6} \text{ M}_{\odot}$, TNG50: $8.94 \pm 4.09 \times 10^{6} \text{ M}_{\odot}$). Interestingly, BHs of AGN galaxies are more massive than non-AGN in TNG100 by roughly one dex, but the opposite is true for TNG50. This is also reflected in the accretion rates where BH in TNG100 on average have $\sim 1.6$ times that of TNG50, but non-AGN in TNG50 have higher accretion rates than non-AGN in TNG100.

Regarding gas densities, there are few things to note. TNG100 and TNG50 follow similar trends with AGN having higher density gas reservoirs than non-AGNs. The similar distributions of BH densities between the two simulations suggest that resolution is unlikely to strongly impact the results. Noteworthy is a small non-AGN population of both TNG100 and TNG50 galaxies with densities around $\log \rho = 3$, i.e a population with little-to-no gas. In fact, out of 548 ($\sim 8.1$ per cent of all dwarfs, $\sim 18.8$ of NOT dwarfs) subhalos with $\log \rho \leq 5$, 487 of them do not have any gas cells associated with them in the group catalogue and 546 (i.e all but two) are considered red ($u-r \geq 2$, red galaxies are discussed further in Section \ref{sec:dis:controlPurity}). This population does not exist in Illustris-1 suggesting that differences in subhalo identification parameters and/or gas physics result in this population.

\subsection{Time since last merger}
\label{sec:TSLM}
\begin{figure}
	\centering
	\includegraphics[width=0.95\linewidth]{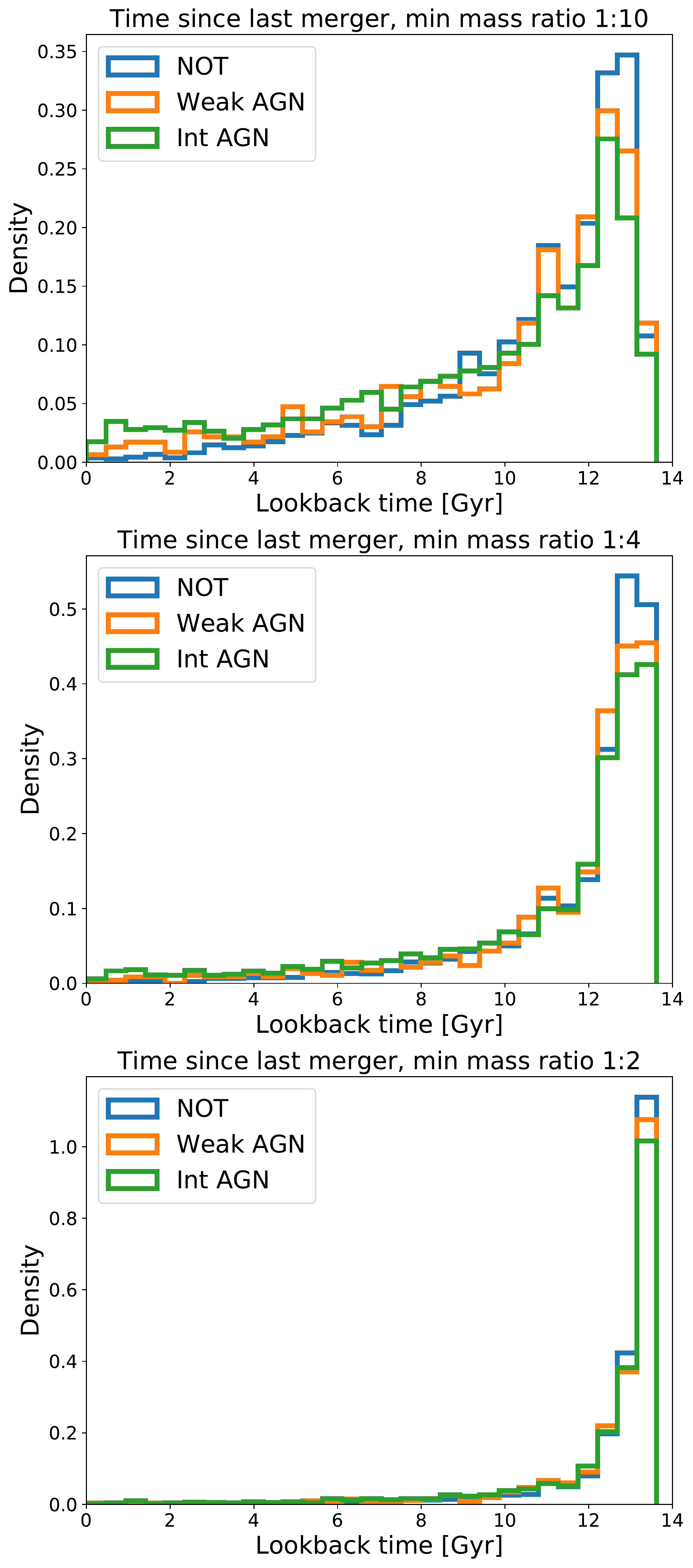}
	\caption{Time since last merger histogram for three selected subsamples: NOT -- low mass galaxies with a black hole but no AGN activity in blue (Section \ref{sec:dwarfSelection}), weak AGN galaxies in orange (Section \ref{sec:AGNSelection}), and intermediate AGN galaxies in green (also Section \ref{sec:AGNSelection}). Galaxies with no mergers have a TSLM equal to the age of the universe}
	\label{fig:TSLMHist}
\end{figure}

The time since last merger (TSLM) is a measure to see if there is a time lag between current AGN activity and a past merger event. While this can already be inferred from morphological disturbances \citep[e.g][]{Ellison2019} or post-star burst spectra \citep[][]{Pawlik2019}, the TSLM method provides a channel in which simulations can explore this connection, too. 


This is done by following the main progenitor branch (MPB) to find the snapshots at which mergers that exceed a merger mass threshold using stellar masses. Three different minimum mass ratios are selected for analysis: 1:10, 1:4, and 1:2. The highest snapshot number is the one chosen as the TSLM. 



The TSLM distributions for different samples can be seen in Figure \ref{fig:TSLMHist} and are then compared against each other using the KS-testing described in Section \ref{sec:ks-testing}.

Since this study is working near the resolution limit, it is vital to ensure that the chosen galaxies are not only of good quality at $z=0$ but also at earlier times. Most (89.5 per cent) have a MPB down to snapshot 5 ($z=9.39$, lookback time 13.286 Gyr) with a tail that includes the rest extending to snapshot 15 ($z=5.53$, lookback time 12.767 Gyr).

\begin{figure}
	\centering
	\includegraphics[width=\linewidth]{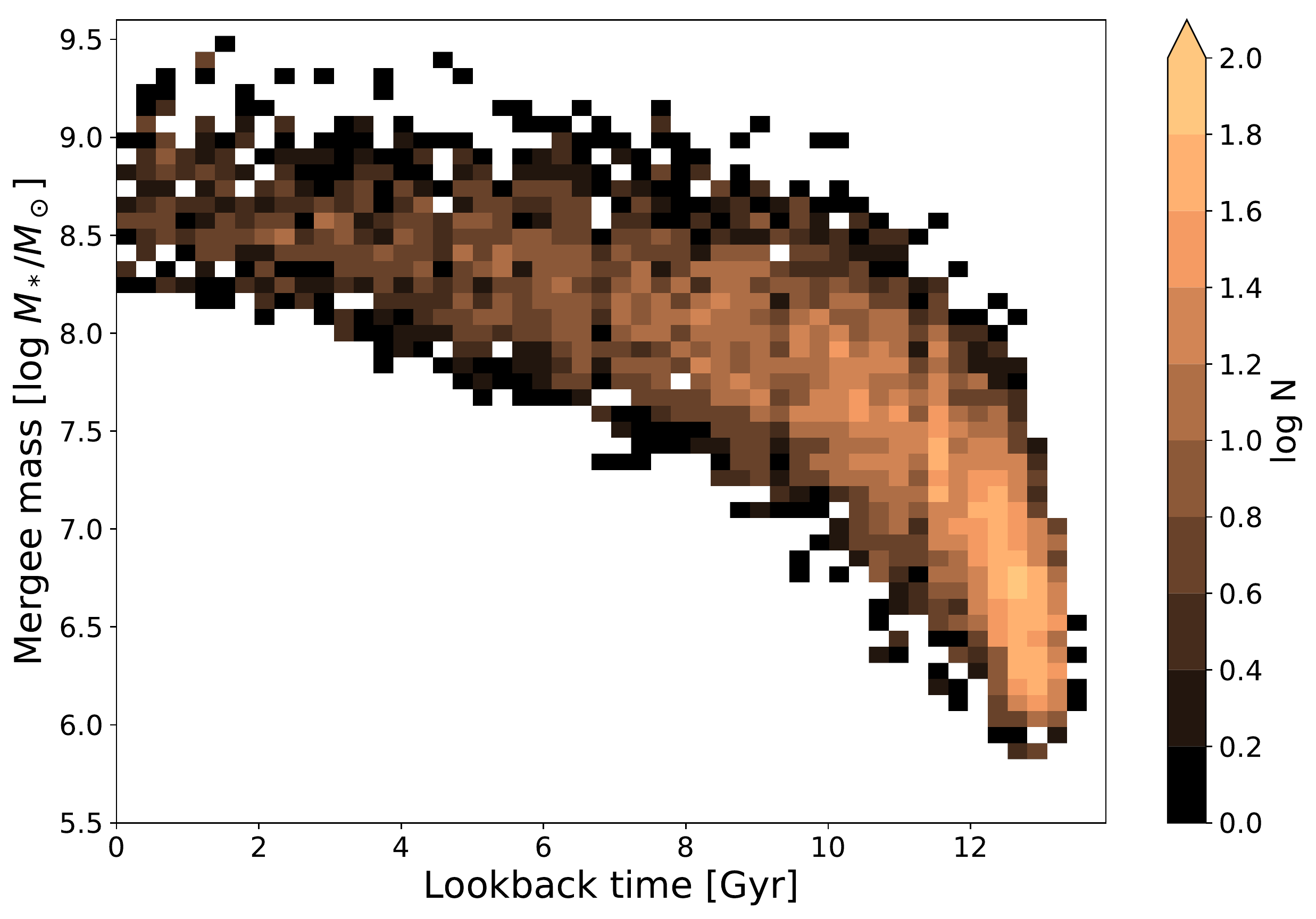}
	\caption{Time since last merger (1:10 mass ratio merger) versus merger stellar mass 2D histogram.  }
	\label{fig:lookbackvsmass}
\end{figure}

This means that there is significant incompleteness at the very early times of the universe. This is further complicated by the fact that the galaxies were less massive earlier which means that e.g a 1:10 merger mass ratio would require a similarly smaller merger mass in order to be counted. This can be seen from Figure \ref{fig:lookbackvsmass} where the majority of the merging galaxies are happening 10 Gyrs ago mostly have a stellar mass of $10^{6.5}-10^{7.0} M_{\odot}$. This means that the merging galaxies consisted of only 10s of particles. Merging galaxies only consistantly reach reasonable resolution ($\sim 100$ star particles, around $3 \times 10^8 M_{\odot}$) at a lookback time of 6 Gyr. As such, any comments on TSLM with lookback times larger than 6 Gyr have uncertainties due to resolution.

\subsection{Distance to 10th nearest neighbour}
\begin{figure*}
	\centering
	\includegraphics[width=\linewidth]{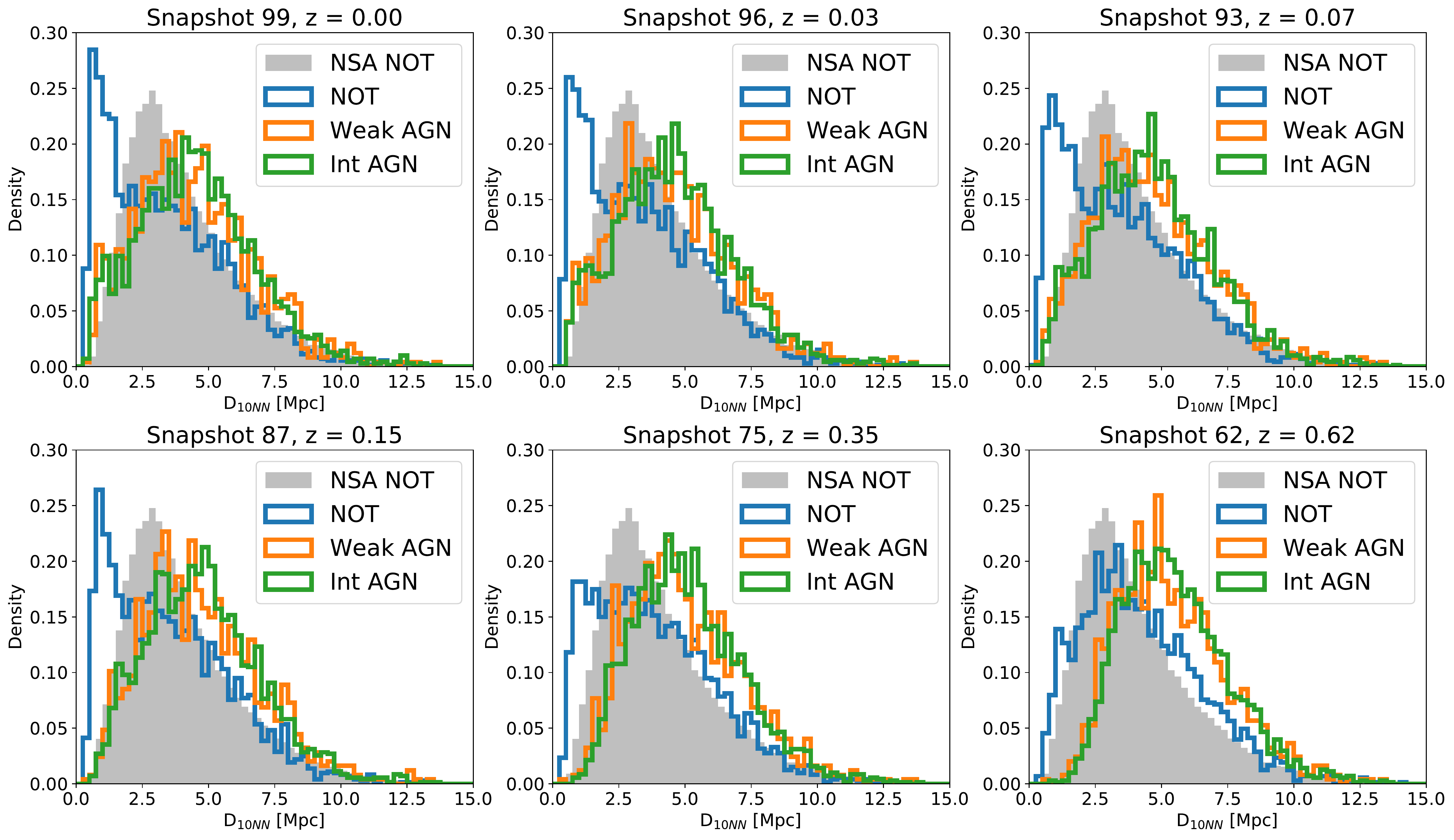}
	\caption{Distance to 10th nearest neighbour histogram for selected snapshots. Blue is non-AGN galaxies, orange is weak AGN while green is intermediate. Strong AGNs are not included as this sample size is small. The grey background histogram is observational data from the NASA-Sloan Atlas (M$_{*} \leq 3\times 10^9$ M$_{\odot}$, $z \leq 0.055$). The snapshots (from high to low) roughly corresponds to lookback times of 0.00, 0.48, 1.00, 1.98, 3.97, and 6.01 Gyr.}
	\label{fig:distHist}
\end{figure*}	

Another way the environment is quantified is by the 3D distance to a galaxy's 10th nearest neighbour, $D_{10}$. This method is used to describe the density of the local environment and was also used in \citet{Kristensen2020}. However, the line-of-sight distance in observations is calculated from redshift and is therefore not the 'true' distance like in simulations. While other types of environmental measures also exist \citep[for a review, see][]{Muldrew2012}, this approach is used because it is a better measure of the local density rather than cluster density and it makes comparisons to previous observational work easier.  While \citet{Haas2012} mention that $N^{th}-$neighbour measurements anticorrelates with halo mass, this anticorrelation is weak for $N=10$ seeing that typical halo masses for low mass galaxies in this sample is $10^{11} \text{ M}_{\odot}$. Therefore, this environment measure probes something else than just halo mass. 

For each galaxy in the sample, the distances to all other 'real' galaxies (defined as non-zero DM mass and above the minimum stellar mass of $10^9 M_{\odot}$) are calculated and then placed in ascending order. Observationally, no restrictions are placed on neighbour galaxies, which may result in more valid neighbours and thus lower $D10$. The 10th element is then chosen and this distance is then chosen as the subject galaxy's distance to its 10th nearest neighbour. 

This measurement is also performed for current AGN galaxies' past environments  $\sim$ 0.5, 1.0, 2.0, 4.0, and 6.0 Gyr ago (i.e at snapshot 96, 93, 87, 75, and 62 (Illustris: 132, 129, 123, 110, 97), respectively). This is done by finding their past progenitor following the main progenitor branch and then repeating the steps described in the previous paragraph.

As with TSLM, the distributions of different subsamples are then compared against each other using the KS-testing method in Section \ref{sec:ks-testing} and the distributions can be seen in Figure \ref{fig:distHist}.

\subsection{Kolmogorov-Smirnov testing}
\label{sec:ks-testing}
The different subsamples are compared against each other using a Monte Carlo approach to 2-sample Kolmogorov-Smirnov (KS) testing. KS testing itself examines the null hypothesis that a sample is drawn from a reference distribution. In the case of a 2-sample KS test, it tests whether the two samples differ (i.e the null hypothesis is whether the two distributions are drawn from the same reference sample). 

The implementation of the test in this study follows that of \citet{Penny2018} and \citet{Kristensen2020} but slightly tweaked. First, a test measure is selected (e.g 10th nearest neighbour at snapshot 99). Then, a subject sample is selected (e.g weak AGN). A random sampling with replacement of $N$ elements
is then performed, and this new sample is then the final subject sample that is to be compared. Next, for each element in the final subject sample, a match galaxy is found in the reference sample (e.g strong AGN). The elements are matched in mass ($\pm 20\%$) and $u-r$ colour ($\pm 0.4$). In the case of multiple matches, a random galaxy is selected and added to the final reference sample while no matches removes the subject galaxy from the final subject sample. The final subject sample and final reference sample are then compared to each other using a 2-sample KS-test. 

This process is repeated 100 times and an average is calculated. This is then repeated 10 times, and the average these 10 tests are then found and subsequently plotted. The error is calculated by finding the average of the standard deviations from the 10 tests and divided by $\sqrt{10}$.

A sampling size of 500 is chosen as the primary sampling size because of its resemblance to observational data as well as to avoid under- and oversampling. However, since there only are 54 strong AGN, any comparisons involving this sample, the sampling size is set to 54. Observational data do permit a larger sample size (see a discussion of this in Section \ref{sec:res:KS_sampSize} and Section \ref{sec:optimalSampleSize}), and testing with sample sizes of N=1\,000, 1\,500, and 2\,000 is also performed, although the size of the weak AGN is only 988 resulting in an upper sampling limit of 988 when comparisons involving weak AGN are done. However, any difference between distributions found only with a higher sampling size means that the difference is less pronounced than with a smaller sample. Therefore, tests with $p\leq 0.05$ and large sampling sizes are only used to infer trends rather than reject the null hypothesis of the two distributions belonging to the same parent distribution.

4 sampling sizes, 5 different samples, and 9 different tests (3 TSLM + 6 $D_{10}$) yield 900 p-values. These are fully plotted in Appendix~\ref{app:ks} in Figures \ref{fig:ks-testsTSLM}, ~\ref{fig:ks-testsNN10_p1}, and ~\ref{fig:ks-testsNN10_p2}). However, significant results are summarised in Table \ref{tab:significantKS} which shows which sampling size the different tests reach significant levels.

\section{Results}
\label{sec:results}

This section contains the results of the KS-testing for the different measures. One thing to keep in mind is that when comparing two subsamples, it is important which subsample is used as subject sample and which as a reference sample. For example for $D_{10}$ distributions, having the NOT subsample as a subject sample and comparing it to the weak AGN subsample, the null hypothesis (i.e that the distributions of the two subsamples belong to the same parent distribution) is rejected. However, the p-value does not reach the threshold when keeping the AGN subsample as subject sample and using the NOT as the reference sample. The reason behind this can be inferred from Figure \ref{fig:distHist-matched} which shows how the non-AGN and weak AGN $D_{10}$ distributions change whether they are subject or reference samples.

While this may not be intuitive at first, it is not surprising. The NOT galaxies have a quite diverse range of masses and colours so that AGN galaxies have a large catalogue to find partners from. 
A larger match catalogue will smooth out the cumulative distribution of the reference sample and thus decrease the average distance between the two distributions. 

Comparisons that reach $p\leq0.05$ with a sampling size of 500 will be called significant while comparisons that reach the threshold at larger sampling sizes are called trends. While these sampling sizes are smaller than fully allowable (i.e not oversampling neither subject nor reference sample), such large sampling sizes are not achievable with current observational data. Having several different sampling sizes allows for different observational surveys to find the closest matching sampling size. A more detailed discussion of this will be given in Section \ref{sec:res:KS_sampSize} and Section \ref{sec:optimalSampleSize}. A table overview of simulation comparisons can be found in Table~\ref{tab:KSSims}

\subsection{On time since last merger}
\label{sec:res:TSLM}
The p-value for 1:10 mass ratio merger in tests with NOT and Int both reach the threshold of 0.05 regardless of which is subject and reference sample. Furthermore, this also holds for 1:4 mass merger ratio of NOT and Int, although it is only within error. Lastly, there is a significant difference within error between Int and weak AGN in 1:10 and a trend between them in the 1:4 case. These results also appear in TNG50-1 and Illustris-1 and will be examined further in Section \ref{sec:res:tng50-ill}.

Interpreting these results show that a merger of at least a 1:10 ratio has happened more recently, on average, in an intermediately active dwarf galaxy than in a non-AGN dwarf galaxy. While this does not establish a causal link (i.e that the merger events triggered the AGN activity), it is a statistically significant difference. \citet{Ellison2019} showed that nearly 60 per cent of mid-IR AGN hosts showed signs of visual disturbances and were either interacting with a close companion or in a post-merger phase with the latter contributing the most. The difference between AGN and non-AGN galaxies in this study is an excess of AGN galaxies with an at least 1:10 merger in the last 0-10 Gyr. 8.4 per cent of Int AGN has had a 1:10 mass ratio merger within the last 3 Gyr compared to only 1.6 of NOT galaxies. Roughly half (55.7 per cent) of Int AGN has not had a merger within the last 10 Gyr, while this number is 71.3 per cent for NOT.

Assuming that most tracers of past merger activity is gone after 1-2 Gyr \citep{ElicheMoral2018}, it is not unlikely that many of the AGN galaxies still retain some merger tracers -- in line with the findings of \citet{Ellison2019}. However, the majority of present day AGN have not had a recent merger within the last 6 Gyr, by which any tracers of a merger most likely are been long gone. This will be discussed further in Section \ref{sec:dis:mergers}.

\subsection{Current and past environments}
\label{sec:res:environments}

\begin{figure}
	\centering
	\includegraphics[width=\linewidth]{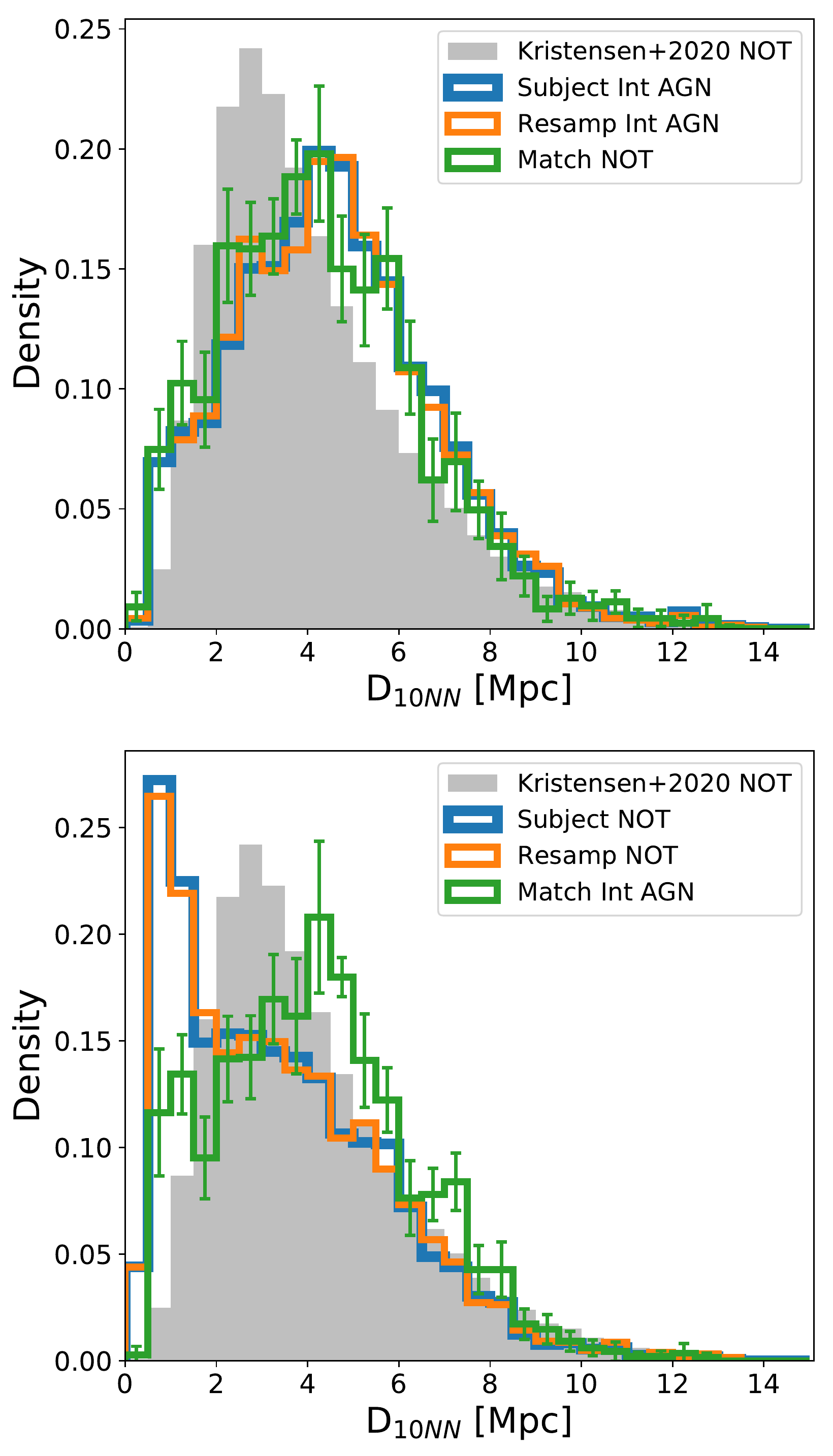}
	\caption{Distance to 10th nearest neighbour histogram with subject samples and their matched reference sample. Blue is the original subject sample, orange is the resampled subject sample while green is the matched reference sample. Top: Int AGN as subject sample and NOT as reference. Bottom: NOT as subject and Int AGN as reference. Errorbars are calculated as the spread of the averages in each bin of 100 resampling runs with a sampling size of 500. The peak at small distances (between 0.5 to 2 Mpc) for NOT disappears when it is used as a reference sample for AGN samples.}
	\label{fig:distHist-matched}
\end{figure}

\begin{figure}
	\centering
	\includegraphics[width=\linewidth]{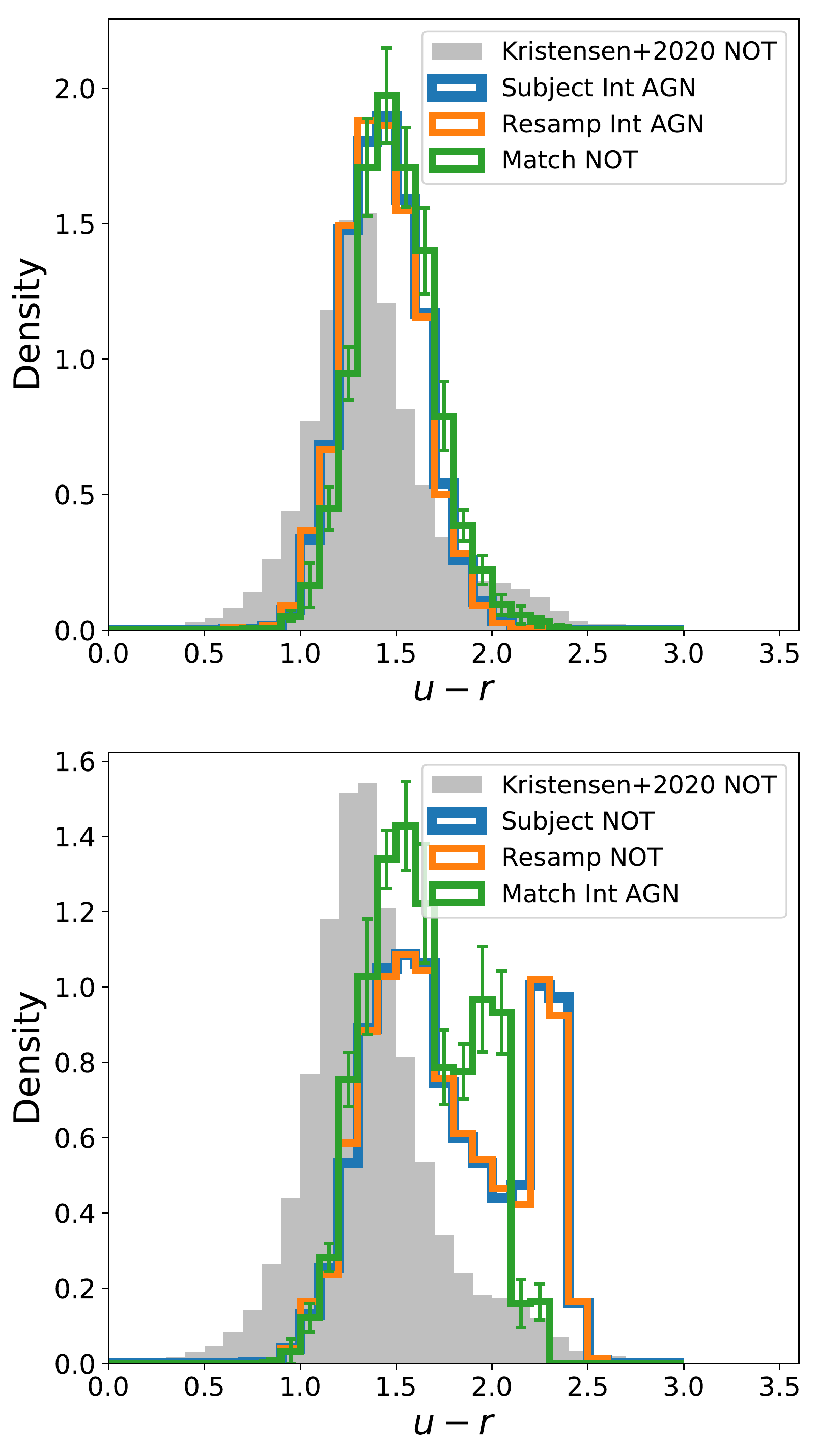}
	\caption{Colour histogram with subject samples and their matched reference sample - like Figure \ref{fig:distHist-matched}. Top plot is using Int AGN as subject sample versus non-AGN as reference sample. Bottom plot is in reverse. Errorbars are calculated as the spread of the averages in each bin of 100 resampling runs with a sampling size of 500. Few red ($u-r \geq 2.0$) NOT galaxies are selected when using AGN samples as subject samples.}
	\label{fig:colourHist-matched}
\end{figure}
Similar across all times is that choosing non-AGN as subject sample and weak or  intermediate intensity AGN as reference sample, the p-value from the results of the KS-tests dips below 0.05. The strong intensity AGN sample do not cross this threshold due to the small sample size. In the inverse situation with the weak and intermediate AGN as subject sample and non-AGN as reference, the p-values do not reach the threshold except at snapshot 62, but do show a trend (using a sampling size of 1000 and above). The similarity between a matched NOT to an Int AGN sample can be seen in Figure \ref{fig:distHist-matched}. These results are similar across both TNG simulations but not Illustris-1.

Worth noticing is the $u-r$ colour distribution of the two samples (see Figure \ref{fig:colourHist-matched}) where both non-AGN and AGN galaxies have a peak around $u-r = 1.5$. However, the non-AGN sample has an additional peak near $u-r = 2.2$ suggesting that a significant sub-population of non-AGN galaxies are very red. Using an AGN sample as subject sample, very few of these red non-AGN galaxies are selected for the reference sample and the resultant $D_{10}$ of the non-AGN reference sample does not have as large a peak of galaxies at $0.5 \text{ Mpc} \lesssim D_{10} \lesssim 2 \text{ Mpc}$. This suggests that this very red sub-population of non-AGN galaxies reside in dense environments. Conversely, very few AGN galaxies reside in these environments. This population and its effect on the results is discussed further in Section \ref{sec:dis:controlPurity}.

\subsection{Sampling size}
\label{sec:res:KS_sampSize}

\begin{table}
    \caption{Summary tables of smallest sampling sizes from KS results.}
    \label{tab:significantKS}
    \centering
    \begin{tabular}{llccc}
        \hline \hline
        {Subject} & {Reference} & {1:10}&{1:4}&{1:2}\\
        \hline
        
        All & NOT & $(\vartriangle)$ & &  \\
        & Weak AGN & & &  \\
        & Int AGN & $(\vartriangle)$ & $(\blacktriangle)$ &   \\
        
        NOT & All & $(\bullet)$ &  &  \\
        & Weak AGN & $\vartriangle $ & $(\vartriangle)$ &   \\
        & Int AGN & $\circ$ & $(\circ)$ & $(\vartriangle)$  \\
        
        Weak AGN & All & & & \\
        & NOT & $(\blacktriangle)$ &  &   \\
        & Int AGN & $(\circ)$ & $\vartriangle$  &  \\
        
        Int AGN & All & $(\vartriangle)$ &  &  \\
        & NOT & $\circ$ & $(\circ)$ & $(\vartriangle)$  \\
        & Weak AGN & $(\circ)$  & $\vartriangle$ &   \\
        
    \end{tabular}
    \begin{tabular}{llcccccc}
        \hline
        {Subject} & {Reference} & {99}&{96}&{93}&{87}&{75}&{62} \\
        \hline
        
        All & NOT & & & & & &  \\
        & Weak & $(\circ)$ & $(\circ)$ & $(\circ)$ & $(\circ)$ & $(\circ)$ & $(\circ)$  \\
        & Int & $\circ$ & $(\circ)$ & $(\circ)$ & $(\circ)$ & $(\circ)$ & $(\circ)$  \\
        
        NOT & All & $(\bullet)$ & $(\bullet)$ & $\bullet$ & $\bullet$ & $(\vartriangle)$ & $\vartriangle$  \\
        & Weak  & $\circ$ & $\circ$ & $\circ$ & $\circ$ & $\circ$ & $\circ$  \\
        & Int  & $\circ$ & $\circ$ & $\circ$ & $\circ$ & $\circ$ & $\circ$  \\
        
        Weak  & All &  &  &  &  &  &  \\
        & NOT & $(\vartriangle)$ & $(\vartriangle)$& $(\vartriangle)$ & $\vartriangle$ & $\vartriangle$ & $(\circ)$  \\
        & Int &  &  &  &  &  &  \\
        
        Int & All & & &  \\
        & NOT & $\vartriangle$ & $\vartriangle$ & $\vartriangle$ & $\vartriangle$ & $\vartriangle$ & $(\circ)$  \\
        & Weak &  & &  &  & &   \\
        \hline
    \end{tabular}
    \tablecomments{Summary tables of which smallest sampling sizes KS results are significant at for TSLM for different miniumum merger mass ratios (top table) and $D_{10}$ at different snapshots (bottom table).  Open dots, open triangles, filled dots, and filled triangles represent sampling sizes of 500, 1000, 1500, and 2000, respectively. Symbols in parenthesis indicates that the $p\leq 0.05$ is reached within error. For a full plots, see Figures \ref{fig:ks-testsTSLM}, ~\ref{fig:ks-testsNN10_p1}, and ~\ref{fig:ks-testsNN10_p2}. Strong AGN are not included since no test reaches the threshold. Tests with weak AGN are limited to a maximum sampling size of 988.}
\end{table}

As mentioned in Section \ref{sec:results}, sampling size has a direct influence on the p-value. This section will motivate that while some samples do not seem to be statistically different (i.e reach $p\leq 0.05$), a trend can still be inferred.
It comes as no surprise that the p-value of a KS-test is dependant on the sampling size since the level at which the null hypothesis can be rejected scales with sample size. More specifically:
\begin{equation}
D_{n,m} > c(\alpha) \sqrt{(n + m)(n \cdot m)^{-1}} ,
\end{equation}
where $D_{n,m}$ is the maximum distance between the cumulative probability of the two distributions, $\alpha$ is the threshold at which to the null hypothesis is rejected, $c(\alpha) = 1.358$ for $\alpha=0.05$, and $n$ and $m$ are the sample sizes. What this means for the interpretation of the above results is that some of the tests that did not yield $p < 0.05$ can reach that threshold given a larger sample size (e.g subject sample weak AGN vs NOT as reference sample with a sample size of 1000). 

Table~\ref{tab:significantKS} shows a summary of the KS tests that reach $p\leq0.05$ at what sampling size. For TSLM, few trends appear for 1:2 mass ratio mergers -- only when using the largest sampling size. A similar pattern can be seen for the 1:4 mass ratio, although the trends are found for the same comparisons samples (NOT vs Int AGN, with both as both subject and reference sample) at a lower sampling size. These two comparisons ultimately reach a significant level at 1:10 mass ratio mergers within error.

For $D_{10}$, the only significant distribution is NOT as subject sample and AGN (both weak and intermediate) as reference samples. As subject sample, weak and intermediate AGN tend to differ at all snapshots from NOT as a reference sample at higher sampling sizes. Ultimately, the usefulness of inferring trends is to estimate the robustness of the tests and to indicate which comparisons are worth further looking into.

\subsection{TNG50-1 and Illustris-1}
\label{sec:res:tng50-ill}
The KS-testing suite with a sampling size of 500 has also been performed on TNG50-1 and Illustris-1. However, the sampling size for the KS-tests never reach a sampling size of 500 since the sizes of the different populations are all below 500. As mentioned in Section \ref{sec:ks-testing}, the sampling size is scaled to the smallest subsample size (e.g in TNG50-1, comparing NOT (size: 417) vs intermediate AGN (size: 337), the sampling size will be 337. 

First, both TNG simulations yield a similar distribution of NOT, weak, intermediate and strong AGNs, although more AGNs are considered weak AGN in TNG50 than in TNG100. The overall percentage of dwarf galaxies considered AGN is the same though (TNG100: $\sim 57$ per cent, TNG50: $\sim 58$ per cent). In Illustris-1, though, the fraction is considerably smaller ($\sim 5$ per cent). This is not surprising considering the different BH seeding and physics between the two simulations. Despite this, results are consistent between simulations, as will be described below.

\begin{table*}
    \caption{Comparison of significant KS results between different simulations.}
    
    \label{tab:KSSims}
    \centering

     \begin{tabular}{lccc}
       \hline \hline
        \textbf{Reference} & \textbf{NOT} & \textbf{Weak AGN} & \textbf{Int AGN} \\
        
        \hline
        \\
        NOT     & -             & 
        \begin{tabular}{|c|c|c|}
        \hline
        $ $ & $ $ & $ $ \\
        \hline
        $\bullet$ & $\bullet$ & \phantom{$\bullet$} \\ \hline
        \end{tabular} 
        &
        \begin{tabular}{|c|c|c|}
        \hline
        $\bullet $ & $ $ & $\bullet $ \\
        \hline
        $\bullet$ & $\bullet$ & $ $ \\ \hline
        \end{tabular}
        \\ \\
        Weak    & 
        \begin{tabular}{|c|c|c|}
        \hline
        \phantom{$\bullet$} & \phantom{$\bullet$} & \phantom{$\bullet$} \\
        \hline
        $ $ & $ $ & \\ \hline
        \end{tabular}
        & - &
        \begin{tabular}{|c|c|c|}
        \hline
        $\circ$ & \phantom{$\bullet$} & \phantom{$\bullet$} \\
        \hline
        $ $ & $ $ & \\ \hline
        \end{tabular} \\ \\
        Int  &
        \begin{tabular}{|c|c|c|}
        \hline
        $\bullet$ & $\circ $ & $\bullet $ \\
        \hline
        $ $ & $ $ & \\ \hline
        \end{tabular}
        &
        \begin{tabular}{|c|c|c|}
        \hline
        $\circ $ & \phantom{$\bullet$} & $\circ$ \\
        \hline
        $ $ & $ $ & \\ \hline
        \end{tabular} 
        & - \\ \\
        \hline
    \end{tabular}
    \\
    \begin{tabular}{lccc}
        & \tiny{TNG100} & \tiny{TNG50} & \tiny{Illustris} \\
        \cline{2-4}
        \tiny{10:1} &\multicolumn{1}{|c|}{$\bullet$} & \multicolumn{1}{|c|}{$\circ$} & \multicolumn{1}{|c|}{$\bullet$} \\
        \cline{2-4}
        \tiny{$D_{10}$} &\multicolumn{1}{|c|}{\phantom{$\bullet$}} & \multicolumn{1}{|c|}{\phantom{$\bullet$}} & \multicolumn{1}{|c|}{\phantom{$\bullet$}} \\ 
        \cline{2-4}
        \end{tabular} \\
    \tablecomments{Columns denote the reference sample while the rows are for subject samples. Each cell is further subdivided into six cells with the columns being the different simulations (TNG100-1, TNG50-1, and Illustris-1 respectively) while the rows are the 10:1 at the top and $D_{10}$ (z=0) is at the bottom. A labelled subtable is shown below (subject Int, reference NOT). A filled circle indicates that the test reached $p\leq 0.05$ while an open circle indicates $p\leq 0.05$ is reached within error.}
\end{table*}
Regarding time since last merger, all three simulations yield a difference in the 10:1 mass ratio mergers between the NOT and Int samples -- although TNG50 only reaches a significant level within error. However, this should be seen in the light of a low sample size where this KS-test only uses a sampling size of 337 for TNG50 (500 for TNG100). Illustris likewise utilises a smaller sampling size (254), but still manages to reach a significant level. 

Regarding NN10, only TNG simulations find significant differences AGN and NOT samples -- and only with NOT as subject. As described in Section \ref{sec:res:environments}, a red dwarf population is largely responsible for this. This subpopulation does not exist in Illustris-1. However, while simply excluding the red dwarfs (and/or requiring a gas component) in TNG does make the NOT sample resemble the AGN distributions more, KS-tests on the NN10 parameter still yields significant differences. This will be discussed further in Section \ref{sec:dis:controlPurity}

\section{Discussion}
\label{sec:discussion}
Five discussions are included below. First is whether or not recent mergers play a significant role in triggering AGN activity. This is followed up by whether there is a time lag between a past environment and current AGN activity. Two more technical discussions ensue where the black hole requirement is first and goes into the details of what effect this requirement has on the sample. Second technical discussion is about whether the dwarf galaxy selection is sufficiently restrained (i.e are the dwarf galaxies found real or are they artifacts of the simulation). Lastly is a discussion about the KS sampling size, and whether this study has used the optimal sampling size or not.


\subsection{Mergers as a significant trigger channel}
\label{sec:dis:mergers}
TSLM found a difference for a minimum merger mass ratio of 10:1 between AGN and non-AGN galaxies -- especially for intermediate intensity AGN. The differences between the distributions are an over-abundance of AGN galaxies with a merger within the last 4 Gyrs and an under-abundance at 10+ Gyr (see Figure \ref{fig:TSLMHist}). However, it is only 11.2 per cent of intermediate AGN with a TSLM $\leq 4$ Gyr and 3.1 per cent for non-AGN, so it is only a minority of all intermediate AGNs in that belong to that bin. 55.7 per cent of intermediate AGN and 71.3 per cent of non-AGN have a TSLM-value of $\geq 10$ Gyr, which also includes no mergers. The fraction of non-mergers increases with merger mass ratio but the fraction is similar between different samples (fractions for 1:10, 1:4, and 1:2 with formatting as $\text{all}^{\text{NOT}}_{\text{Int}}$: $2.6^{2.6}_{2.6}$, $16.7^{17.6}_{16.0}$, and $45.3^{47.7}_{43.3}$ per cent). For weak AGN, the numbers are 6.9 percent and 65.1 percent, thus showing a similar but weaker trend as intermediate intensity AGN. The fraction of galaxies that has had a merger in its past but not recently (i.e $4  \text{ Gyr}\leq \text{TSLM} \leq 10 \text{ Gyr}$) is 25.6, 28.0, and 33.1 per cent for non-AGN, weak, and intermediate AGN, respectively. 

As mentioned in Section \ref{sec:TSLM}, merger activity further than 6 Gyr ago has to be considered with a grain of salt. Therefore, we pertain ourselves to only a distinction of \textit{recent} activity meaning TSLM $\leq 4$ Gyr and no or unaffected by mergers TSLM $\geq 6$ Gyr.

Interpreting on these numbers, it means that a dwarf galaxy with a recent merger is more likely to host stronger AGN activity, although it is not a requirement since the majority of active galaxies have had longer a longer time since a minor merger. In fact, the median TSLM value of non-, weak, and intermediate  AGN is 11.56, 11.26, and 10.52 Gyr, suggesting that most AGN galaxies are still unaffected by merger activity. So while other factors are in play for triggering most AGN activity in dwarf galaxies, mergers seem to be associated with increased and stronger AGN activity. This is similar to findings in the EAGLE simulation by \citet{McAlpine2020} who found that mergers increase rate of luminous AGN. In our study, the sample size for strong AGN activity is too small to make convincing conclusions. 

An important note to make is that this study only examined the role of mergers. \citet{Martin2020} remark that mergers only drive 20 per cent of morphological disturbances in the NewHorizon cosmological simulations but are instead most often due to interactions that do not result in a merger. Such interactions are not picked up and looked at in this study but may be an important channel for AGN triggering. 

While mergers do not appear to be the most significant trigger channel, it cannot be dismissed. Further and more complex examination of the dynamics of especially intermediate AGN dwarf subhalos may be needed to map out the cause of this increase.

\subsection{Time lag and impact from past environments}
\label{sec:dis:environment}

\begin{figure}
	\centering
	\includegraphics[width=\linewidth]{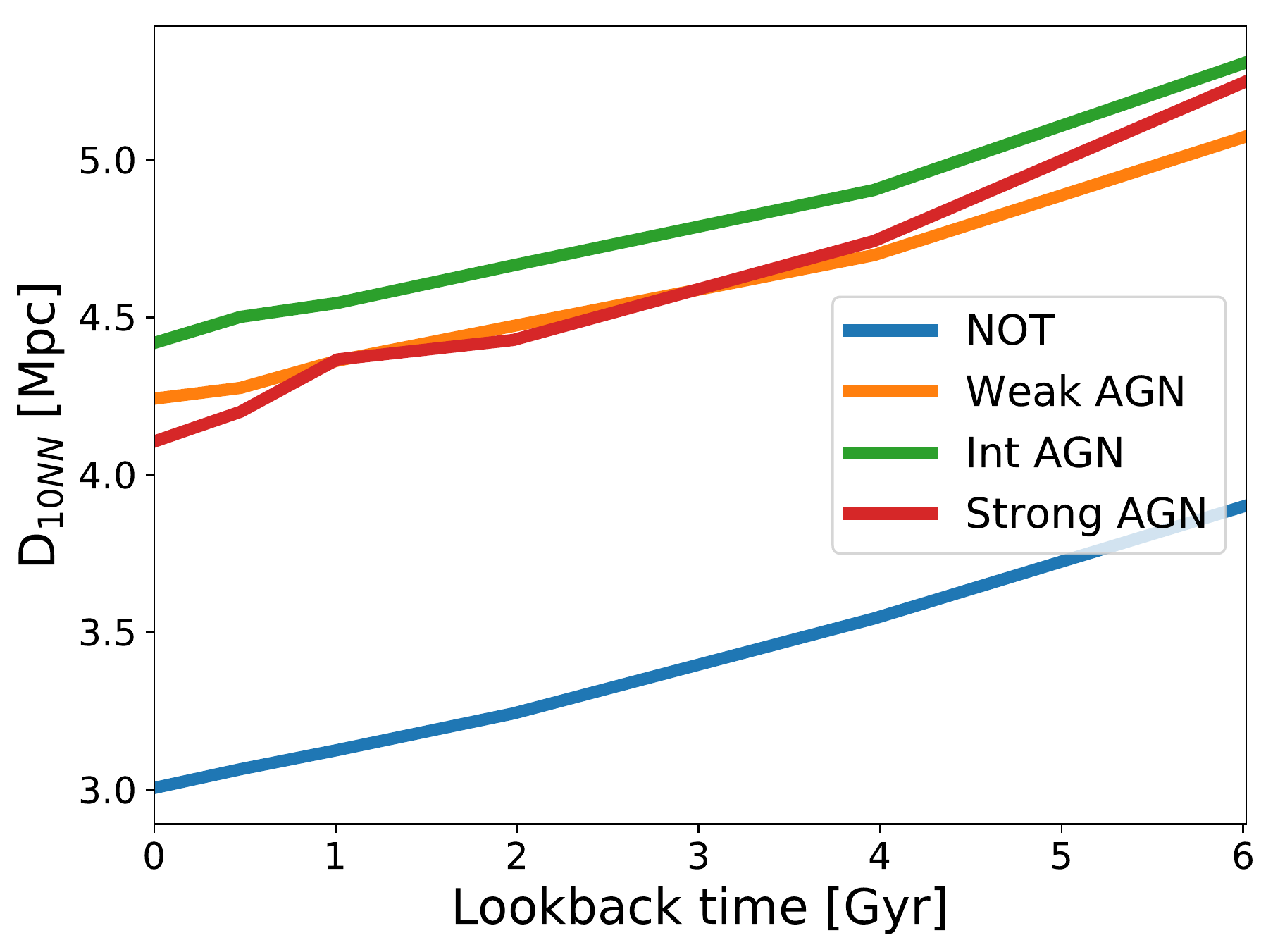}
	\caption{Distance to 10th nearest neighbour evolution. Each line is the median of the $D_{10}$ distribution of different samples at different snapshots.}
	\label{fig:10NNevo}
\end{figure}

A question that is attempted to be answered in this study is whether or not the past environment can trigger or at least may lead to AGN activity further down the road. What is found is that the past environments of both $z=0$ AGN dwarf galaxies and non-AGN galaxies are towards higher $D_{10}$ values (i.e less dense environments, see Figure \ref{fig:10NNevo}) but that they otherwise maintain their differences; the average $D_{10}$ of the subsamples increase similarly, but the difference between their averages remain the same in all snapshots. Similarly, the appearance of the distributions of the subsamples all flatten going back in time, but their overall shape remains the same which means they maintain their differences.

A non-insignificant number of non-AGN galaxies that are red ($u-r\geq 2.0$) are found in dense environments ($D_{10} \leq 2$ Mpc). More specifically, around 31 per cent of non-AGN galaxies are red, and of those 31 per cent, 76 per cent of them are found in dense environments. For blue non-AGN galaxies, only 16 per cent are in dense environments, and for Int AGN galaxies, only 12 per cent are in dense environments. All of this is to say that almost all of the red non-AGN dwarf galaxies are in dense environments while only a few of blue non-AGN and AGN galaxies are in dense environments today.

Furthermore, the red peak (see Figure \ref{fig:colourHist-matched}) is not significantly present for $z=0$ non-AGN galaxies at $z=0.7$ (snapshot 59) with only 1.1 per cent has $u-r\geq 2.0$. This number grows to 4.3, 11.4, 16.1, and 21.7 per cent for $z=0.5, 0.3, 0.2, 0.1$, respectively. Of the red non-AGN dwarf galaxies at $z=0$, 89.3 per cent of them are already in dense environments at $z=0.7-0.6$. This is in contrast to both $z=0$ AGN and blue non-AGN dwarf galaxies of which only 1.0 and 35.0 per cent are in dense environments at $z=0.7-0.6$, respectively.

Given that the colour is calculated from the stellar particles in the galaxies with a dust attenuation model, a red colour suggests either an old stellar population or strong dust attenuation. For a galaxy in a dense environment that also has a shallow potential well, stripping of its gas reservoirs can quench star formation and thus be left with an aging stellar population. \citet{Sabater2015} suggest from observations that the level of nuclear activity in galaxies depends on the availability of cold gas in their nuclear regions. Applying this explanation for the results of this study, it would mean that dense environments strip dwarf galaxies of their cold gas halting star formation and AGN activity, too. 

As mentioned in Section \ref{sec:data:bh_comparison}, all subhalos with low gas density near the BH have no gas cells associated with them and are red. However, this only accounts for slightly more than half of the red subhalos (548 out of 915 red galaxies). Generally, the red population has fewer gas cells (median red subhalo $n_{\text{gas}} = 0$, red subhalos with gas cells $n_{\text{gas}} = 475$, all dwarfs $n_{\text{gas}} = 8108$). This low count of gas cells can be due to stripping, or some other physical mechanism, or due to the Subfind algorithm, although the red population does not exist in Illustris which uses a similar Subfind algorithm.

Still, this does not answer the question whether circumstances in the past has led to AGN activity now. However, a dense past environment can be a strong indicator of whether or not AGN activity is likely in the future, and if a dwarf galaxy has been in a dense environment in the past $\sim 6$ Gyr, it is unlikely to host AGN activity.


\subsection{Black hole requirement}
\label{sec:BHDiscussion}
\begin{figure}
	\centering
	\includegraphics[width=\linewidth]{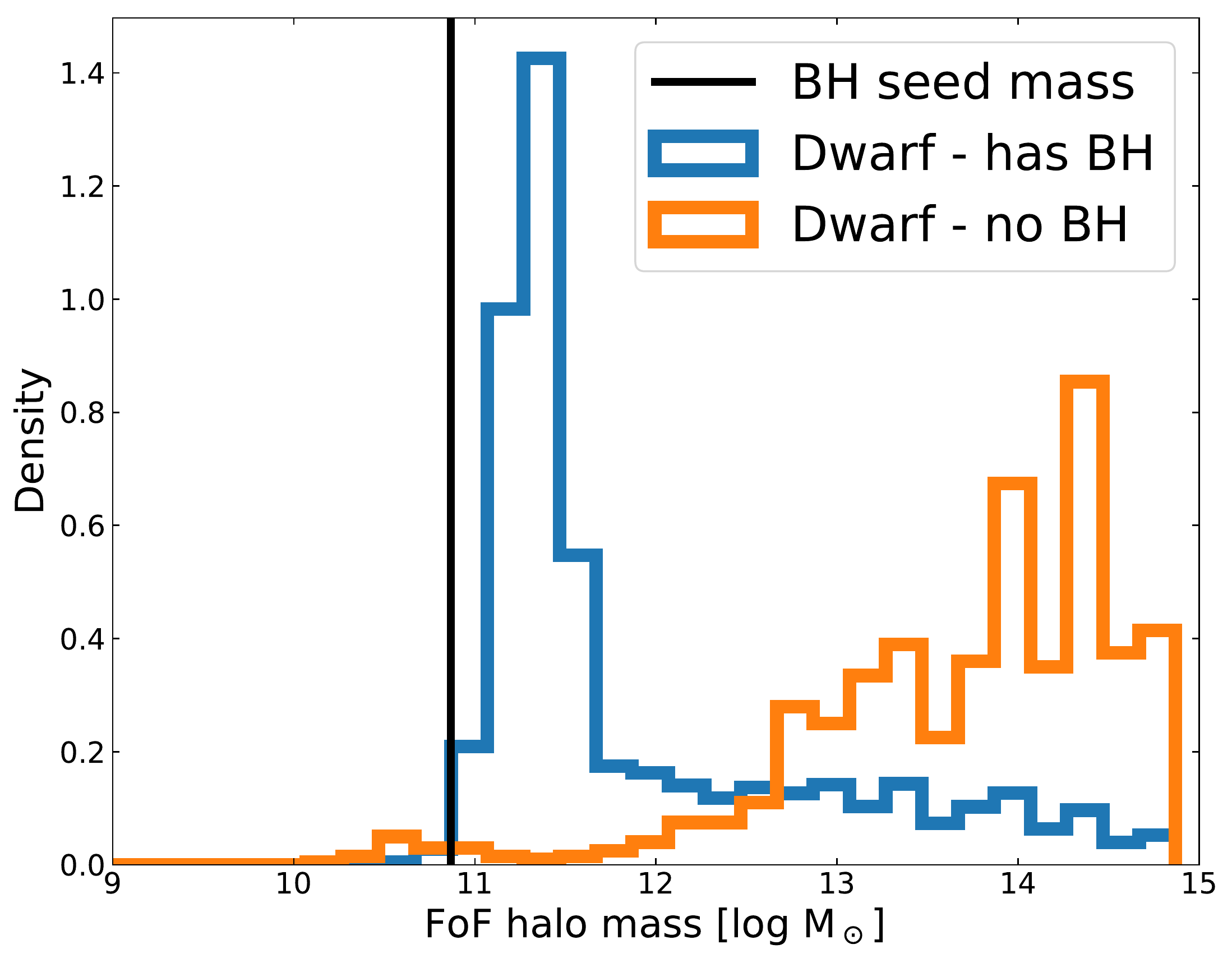}
	\caption{Mass distribution of low mass galaxies with (blue) and without (orange) BH. Additionally, the FoF halo mass threshold ($7.38 \times 10^{10}$ M$_{\odot}$) for when a BH is seeded is shown as a black line. Galaxies with (without) a BH, 0.7 per cent (2.0 per cent) have a lower mass than the seed threshold and 99.3 per cent (98.0 per cent) have a higher mass.}
	\label{fig:massHist_BH-noBH}
\end{figure}

As described in Section \ref{sec:dwarfSelection}, only galaxies with a black hole are included in the dwarf galaxy sample. Due to the way BH seeding works in IllustrisTNG, about an eighth of dwarf galaxies are left without a black hole, which consequently means they will never show up as having AGN emission. This is despite the fact that their real life counterparts may very well host a BH. 

Overall, there are two categories of no-BH dwarf galaxies, which can be inferred from Figure \ref{fig:massHist_BH-noBH}: 1) Those whose FoF halo are below the mass threshold for seeding (roughly 2.0 per cent), and 2) those whose FoF halo is larger than the mass threshold (roughly 98.0 per cent). These two categories will have different cosmological histories. For a minimum stellar mass cut of $10^8 M_{\odot}$, these percentages change to 27.4 per cent below and 72.6 per cent above the seeding threshold, which suggests that the first scenario is more common in lower mass galaxies.

The light FoF halos are assumed to have never reached the FoF halo mass threshold and thus constitute isolated galaxies that have evolved secularly and only had few to no mergers in its past. The dwarf subhalos in massive FoF halos are presumed to similarly never have been able to reach the BH seed mass threshold but whose FoF halo has merged with another halo with either a BH already (and thus are restricted from being seeded a BH) or a more massive subhalo in which the BH would then be seeded in (since if two or more subhalos exist in the same halo, only the most massive subhalo would be seeded a black hole). 

\begin{figure*}
	\centering
	\includegraphics[width=\linewidth]{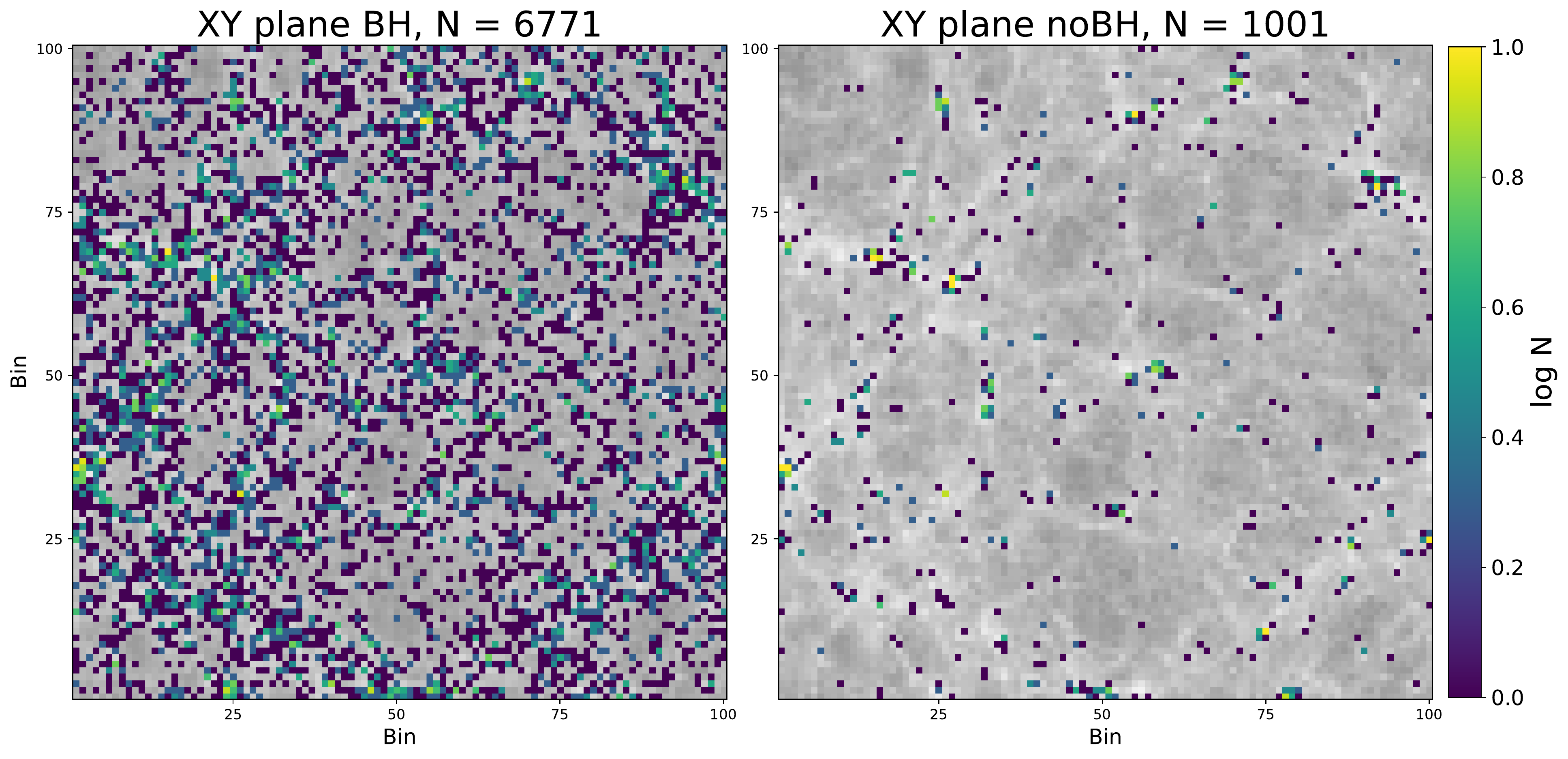}
	\caption{Spatial distribution on the XY plane of low mass galaxies with (left) and without (right) BH. There are 100 bins on each axis and the number of subhalos in each bin is then counted.}
	\label{fig:density_BH-noBH}
\end{figure*}

\begin{figure}
	\centering
	\includegraphics[width=\linewidth]{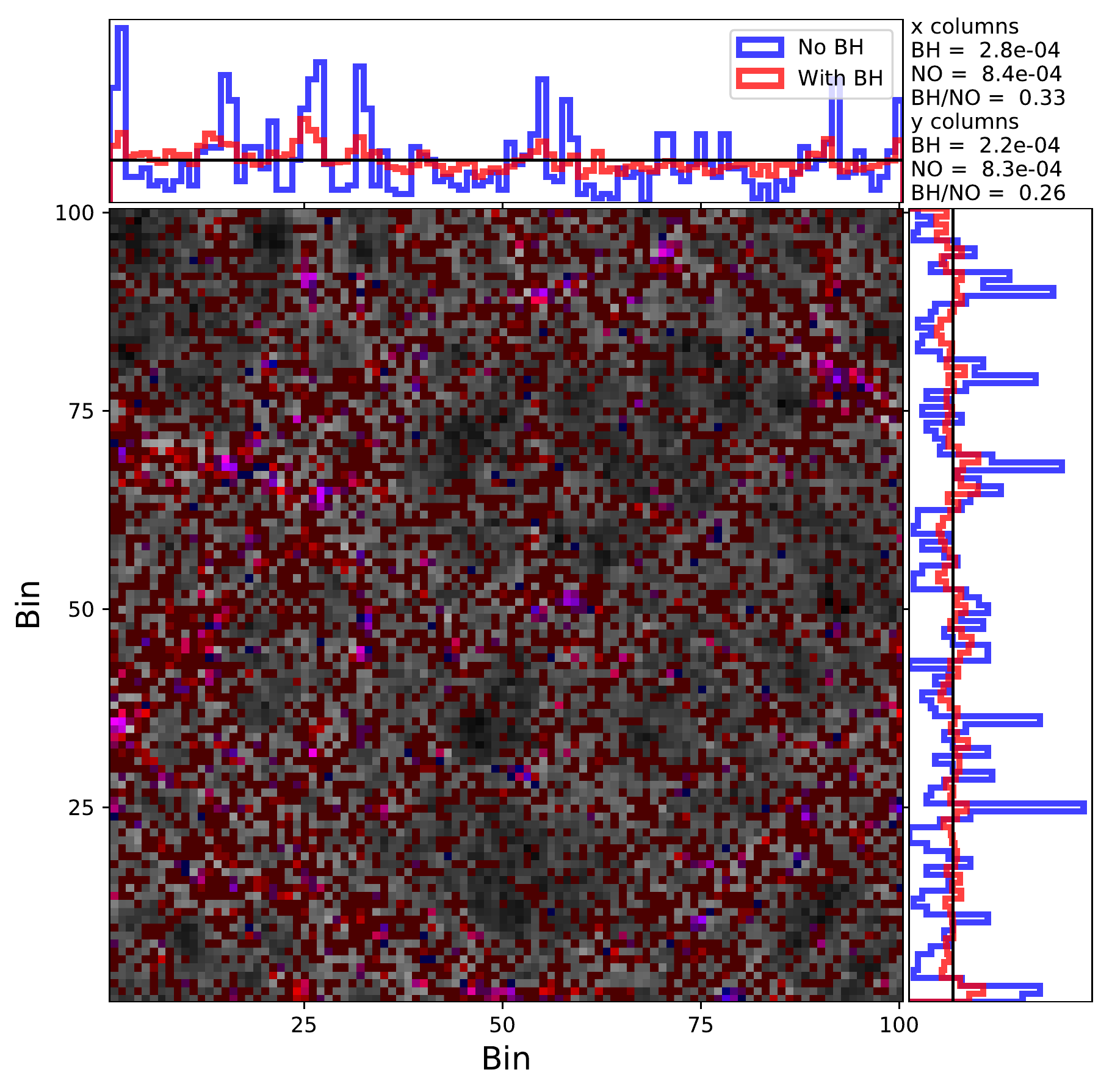}
	\caption{Spatial distribution on the XY plane of low mass galaxies with (red) and without (blue) BH. There are 100 bins on each axis and summed up in a normalised histogram. Each pixel bin is given a colour corresponding to the ratio between the number of BH to no-BH galaxies with more blue meaning a higher number of no-BH galaxies. The black line in the histograms show the average density, i.e the density distribution if all galaxies were spread out evenly. The departure from the this distribution of the BH and no-BH distributions is calculated and shown next to the histograms. The residual of the no-BH distribution is higher for all axes indicating that they clump together moreso than BH galaxies.}
	\label{fig:densityPlot_wHist}
\end{figure}

The number density distribution of the dwarf subhalos with either a BH or no BH supports the idea of the different cosmological histories between the samples. Figure \ref{fig:density_BH-noBH} shows the two samples side by side in a log N density plot, and despite the different sizes of the samples, the no-BH sample is more clustered with multiple spots of log N densities of around 1.0. Few or no such spots appear in the BH sample. The densities are found from binning coordinates in 100 bins and then counting the numbers of subhalos in each bin.This is further quantified in Figure \ref{fig:densityPlot_wHist} where dwarf subhalos with BH have a lower residual from an average spatial distribution compared to dwarfs with no BH.

These considerations show that certain demographics of dwarf subhalos are excluded: The ones that are very isolated and the ones in very dense environments, although the former is negligible consisting of less than 1 per cent of the total population. That means that the data set used in this study is unable to satisfyingly describe extreme scenarios of dwarf subhalos with AGNs and any conclusions are limited to the more moderate population. 

\subsubsection{Bias from exclusion of no-BH subsample}
\label{sec:dis:bhBias}
In order to quantify the bias from this exclusion, the KS tests are run with a sampling size of 500 where the non-black hole galaxies are added to either the non-AGN sample or the intermediate AGN sample. In the modified samples, the no-BH sample constitutes 25.6 per cent and 26.2 per cent in the NOT and Int samples, respectively. The no-BH population by itself is characterised by small $D_{10}$ (and high halo masses) and a similar TSLM distribution as the NOT sample. From these considerations, it is expected that adding the no-BH sample to the NOT sample will amplify the already exisiting difference between NOT and Int regarding both mergers and $D_{10}$ while adding the no-BH sample to Int will lessen the differences. The question is then if this is to a significant degree.

Regarding time since last merger, TSLM, the results stay the same, although adding the no-BH to Int does increase the p-value for 10:1 TSLM to just above 0.05 (with NOT as both subject as reference sample versus the modified Int sample), but the error extends below 0.05. For the 1:4 TSLM, using NOT as subject and modified Int as reference, the threshold is now linger within error. Adding the no-BH to NOT does not change the result, except making the p-value even smaller.

On the distance to the 10th nearest neighbour, $D_{10}$, results are also somewhat as expected but with some indication that the no-BH population is different to both the NOT and Int populations. First off, adding the no-BH sample to the Int sample, the p-value is now below 0.05 for NOT and modified Int as both subject and reference sample (compared to only NOT as subject and Int as reference in normal testing). Adding the no-BH sample to NOT maintains the original results (subject NOT and reference Int $p\leq 0.05$), but having Int as subject and modified NOT as reference now reaches the threshold within error.

The NOT and Int samples are also compared against their modified samples. For TSLM, NOT is able to produce a p-value above the threshold while Int does not (within error). This is as expected since the no-BH TSLM distribution closely resembles the NOT distribution while differing from the Int distribution. For $D_{10}$, neither sample is able to reach above $p=0.05$ when their modified sample is used as subject sample. Furthermore, no tests reach the threshold when both samples are modified (e.g modified NOT vs modified Int), but this can be interpreted as subject no-BH galaxies being able to match with themselves and/or other galaxies within no-BH with whom they share characteristics with. 

Summarily, in the extreme cases where the no-BH subhalos belong to either non-AGN or intermediate AGN samples, the TSLM results remain unchanged (except for 1:4 mergers) although less certain when the Int sample is modified. The picture is muddled regarding $D_{10}$ where the no-BH sample inhabits a different parameter space compared to both NOT and Int samples. The initial results still hold but requires an added complexity to the interpretation of the results; the typical environment of the no-BH sample is very dense ($D_{10} \leq 1.0$ Mpc) and belonging to a very massive halo (M$_{\text{halo}} \geq 13\text{ M}_{\odot}$, and this type of environment is not typically seen among the other samples. Whichever sample the no-BH is added to would introduce a unique environment and may thus yield a significant difference in $D_{10}$ distributions. Ultimately, by not using no-BH subhalos in the main analysis and results makes us unable to gauge the impact of the very dense environments, but even when they are included, the results stay the same. Similar trends are found in TNG50-1 and even Illustris-1 despite its different BH prescription.

\subsection{Are control galaxies properly constrained?}
\label{sec:dis:controlPurity}


The selection of a proper control sample (in this study, it is synonymous with the 'NOT' sample) is important since comparison to this constitutes the basis of the statistical analysis. A biased control sample will give an impression that comparison samples follow different distributions and may lead to interpretations of their environment and past. Section \ref{sec:BHDiscussion} discusses one selection criteria (i.e requiring a black hole) that removes a significant amount ($\sim 13$ per cent) of the low mass galaxy sample. Most of these are in dense environments (see e.g Figures \ref{fig:densityPlot_wHist} and \ref{fig:density_BH-noBH}) thus resulting in the overall distribution moving towards less dense environments/larger $D_{10}$ distances. 

Even after this correction, the control sample still has a second peak in the $D_{10}$ distribution around 0.5-1.5 Mpc. While the KS-testing does reveal that NOT distributions and matched AGN distributions do not follow the same parent distribution, Figure \ref{fig:distHist-matched} shows that this discrepancy can be almost nullified if only the blue NOT galaxies are considered (i.e matched NOT to weak AGN follow similar distributions). Since there are only blue AGNs, a matched reference sample to an AGN sample must also tend towards being blue. Nevertheless, there are two possible scenarios: 1) The control sample is properly chosen which means that dwarf galaxies in dense environments in TNG tend not to develop AGN activity, or 2) the control sample is not sufficiently constricted and thus that the $D_{10}$ measure is biased.

Regarding the first point, several of the selection criteria are in place to avoid biases from technical parts of the simulation. That is not to say that further technical biases do not exist, but since the main contributors to a bias in the control sample have been identified and corrected, we conclude with this caveat in mind that dwarf galaxies in dense environments in TNG100-1 are less likely to develop AGN characteristics -- possible due to a lack of gas. This similarly is the case for TNG50-1. In Illustris-1, though, this population is not present, suggesting either a systematic galaxy definition difference or dwarf galaxy population difference between TNG and Illustris -- maybe due to a difference between stellar and AGN feedback models in Illustris and TNG. 

Observationally, with a similar method, no such trend is found \citep[]{Kristensen2020}, and a double peak in the distance distribution is also not found (i.e one between 0.5-2.0 Mpc and another near 2.5 Mpc, see NSA NOT distribution in Figure \ref{fig:distHist}). This suggests that the control sample is not sufficiently constricted. It may be a manifestation of the missing satellites problem where there actually \textit{is} a concentration of dwarf galaxies in dense neighbourhoods but they are not observable resulting in a missing peak at 0.5-2.0 Mpc in observational data \citep[e.g ][for a Milky Way-Andromeda like system]{Fattahi2020}. 

However, the most recent and highest resolution IllustrisTNG simulation run, the TNG50 \citep[][]{Pillepich2019, Nelson2019a} run, has found that observations and simulations are in good agreement for Milky Way-Andromeda like systems \citep[][]{Engler2021}, down to a stellar mass of $10^7$ M$_{\odot}$. This stellar mass threshold is above the lower mass threshold of this study, so the missing satellites problem seems an unlikely culprit. The  population also exists in the TNG50-1 data, but does not exist in Illustris-1. This indicates that this population is systematic to TNG. If this population does not exist in other simulations, it would suggest that the red dwarf galaxy population in TNG100-1 (and TNG50-1) is of a non-physical origin and should be excluded. Adding to this argument is the work of \citet{Dickey2021} that found an overestimation of the quiescent fraction of isolated dwarf in simulations compared to observations. 

The significance of this red dwarf population on the results is tested by removing dwarf galaxies with $u-r \geq 2.0$ in the NOT sample, which removes 891 dwarf galaxies resulting in a modified NOT sample size of 2\,017. While the peak near 1 Mpc in the $D_{10}$ distribution shrinks, there is still a noticeable plateau between 0.5-3.5 Mpc (which is not present in AGN samples) and KS-testing (500 sampling size) does indeed still show a significant difference in distributions between modified NOT versus Int galaxies, although it changes to be only within error. TSLM results are unchanged. This lends credence to the results from this study -- at least in TNG simulations

\subsection{Optimal KS sampling size}
\label{sec:optimalSampleSize}
There are several considerations when choosing the sampling size. One point is regarding the effect on the KS statistics (see Section \ref{sec:res:KS_sampSize}) with a larger sample size yielding lower p-values. Obviously, if there is a statistical difference between two distributions, then it is desirable that the testing shows this. However, large sample sizes (especially if \textit{oversampling}) may exaggerate small differences that may be due to random error.

Another consideration is the resemblance to observations. In \citet{Kristensen2020}, $\sim 40\,000$ low mass galaxies were used and $\sim 200-4\,000$ of these were classified as AGNs (depending on selection method). The data used was the NASA-Sloan Atlas, which covers around 1/3 of the sky to a very high level of completeness at $z \leq 0.055$. Assuming these numbers are near the current observational limits and that oversampling is not desired, this effectively limits our sample size to $\sim 200-4\,000$.

Although a fixed sample size of 152 was used in \citet{Kristensen2020}, 500 is used in this study since it is well within the observational range described in the previous paragraph. Similarly, it does yield several tests with p-values below 0.05 (e.g, in $D_{10}$ distributions, subject sample NOT vs AGN samples as reference samples, see Section \ref{sec:results}). However, as mentioned in Section \ref{sec:res:KS_sampSize}, increasing the sample size to even 1\,000 provides further comparisons that drop below 0.05, which suggests at the very least that a trend exists in those comparisons. 

\citet{McAlpine2020} remark that both minor ($1:10 \leq M_{1}/M_{2} \leq 1:4$ and major mergers ($M_{1}/M_{2} > 1:4$) play a role in black hole activity (but not significantly in black hole growth) in the EAGLE simulation -- a relation that is also can be inferred in this study with a sample size of more than 1\,000, although not significantly with a sample size of 500. Thus a sample size in this range reproduce results from similar studies.
Observationally, \citet{Ellison2019} similarly find that AGN activity is enhanced in mergers and disturbed systems lending further credence to a sample size in this range.

However, \citet{Shah2020} find no enhancement of AGN activity for close pairs interacting, except for visually identified systems or those that has already coalesced, although they remark that their results are also consistant with low-level AGN enhancement. As discussed in Section \ref{sec:dis:mergers}, mergers do not seem to be a major trigger channel in IllustrisTNG data, but minor enhancement can be interpreted from the results of this paper.


\section{Summary}
\label{sec:summary}
The environments of non-AGN and matched AGN (of both weak and intermediate intensity) galaxies are different from each other with non-AGN prefering denser environments ($D_{10} \leq 2$ Mpc). Environments of non-AGN matched in stellar mass and colour to AGN galaxies are not significantly different to each other.

Around 31 per cent of dwarf galaxies that do not develop AGN characteristics  are red ($u-r \geq 2$)  and 76 percent of those are located in dense environments at $z=0$. Around 6.2 Gyr ago, most were blue (99 per cent) but around half (47 per cent) were already in dense environments (and 89 per cent within $D_{10} \leq 4$ Mpc). However, even ignoring red dwarf galaxies yields a significant difference between environments of AGN and non-AGN galaxies.
This suggests that prolonged exposure to dense environments is not only detrimental to star formation (from the increasing fraction of red galaxies) but also AGN activity -- at least in TNG. 

1:10 mass ratio mergers are to a significant degree different between intermediate intensity AGN and non-AGN galaxies. The difference is primarily recent mergers ($\text{TSLM} \leq 3$ Gyr) and distant/no mergers ($\text{TSLM} \geq 10$ Gyr or no merger) with intermediate intensity AGN having had more recent merger activity than non-AGN. No such difference is seen in other mass ratios, although a 1:4 ratio is following the same trend but is not significant.

This suggests that for a minority (around 8.4 per cent) of $z=0$ intermediate intensity AGN, a small merger can lead to increased AGN activity in dwarf galaxies, although it is not always the case since 1.6 per cent of non-AGN galaxies have also had a recent merger. Observations point in both ways depending on the method employed to determine recent merger history and statistical significance level. 

Lastly, there are caveats working in this mass regime in cosmological simulations. One seventh of the dwarf galaxies are not included as the TNG seeding criterion does not assign a BH to these dwarfs and they mostly belong to very dense environments leaving this environments unexplored. However, the bias from excluding this population is negligible. Also, a population of non-AGN galaxies in dense envrionments is present in TNG runs but not Illustris, suggesting either a systematic galaxy definition difference or dwarf galaxy population difference between TNG and Illustris.

\section*{acknowledgements}
      We acknowledge the support of STFC through the University of Hull Consolidated Grant 
ST/R000840/1, and access to {\sc viper}, the University of Hull High 
Performance Computing Facility.

\begin{appendix} 
\section{Full KS results visualisations} 
This section contains the full page visualisations of all the KS-results for TNG100-1.  
\label{app:ks}
\begin{figure*}
	\centering
	\includegraphics[width=\linewidth]{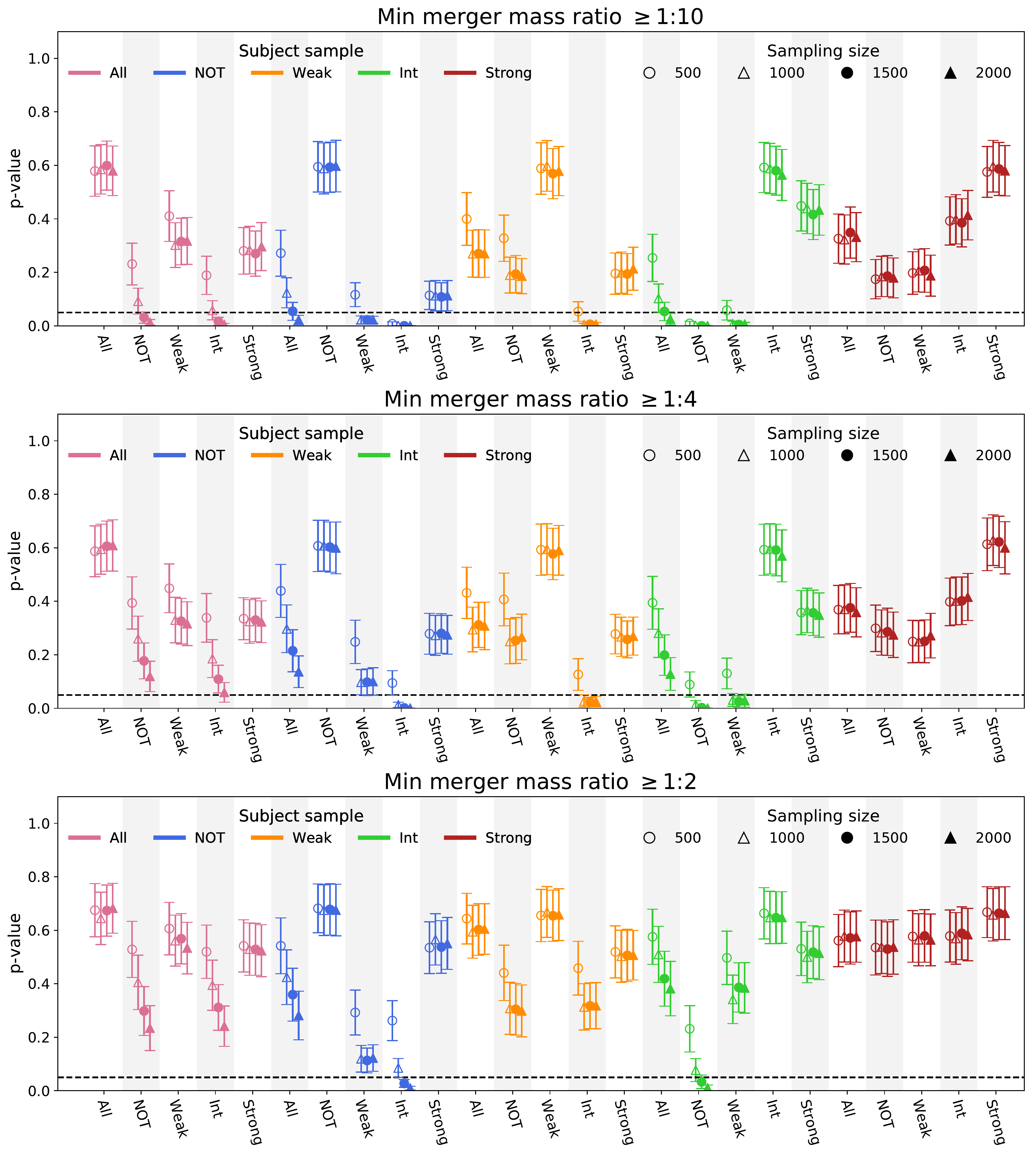}
	\caption{KS-testing results of merger mass ratio. Details on how the p-value and its error is calculated can be found in Section \ref{sec:ks-testing}. Colour indicates what the subject sample is with violet being all dwarf galaxies (i.e NOT+Weak+Int+Strong), blue being non-AGN, orange is weak AGN, green is intermediate AGN, and red is strong AGN. On the x-axis is the reference samples with the marker style indicating sample size. Background shading indicates a group of data points with the same subject and reference sample. For example, if you are to look up what the p-value is for non-AGN as subject and weak AGN as reference using a sampling size of 500, it is found as the orange open circle at the 8th tick mark on the x-axis.}
	\label{fig:ks-testsTSLM}
\end{figure*}
\begin{figure*}
	\centering
	\includegraphics[width=\linewidth]{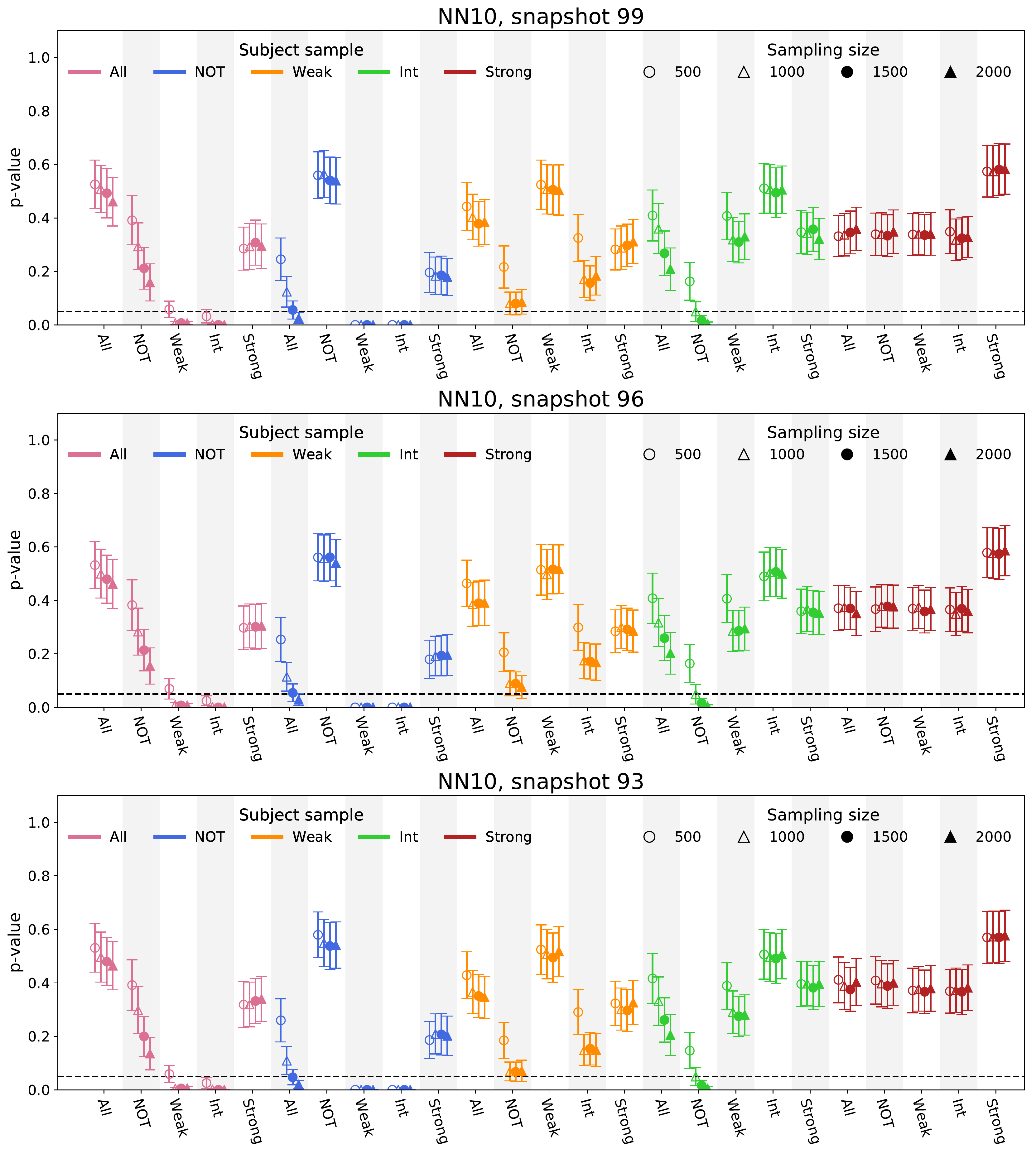}
	\caption{Same as Figure \ref{fig:ks-testsTSLM} but for $D_{10}$ for snapshot 99, 96, and 93}
	\label{fig:ks-testsNN10_p1}
\end{figure*}
\begin{figure*}
	\centering
	\includegraphics[width=\linewidth]{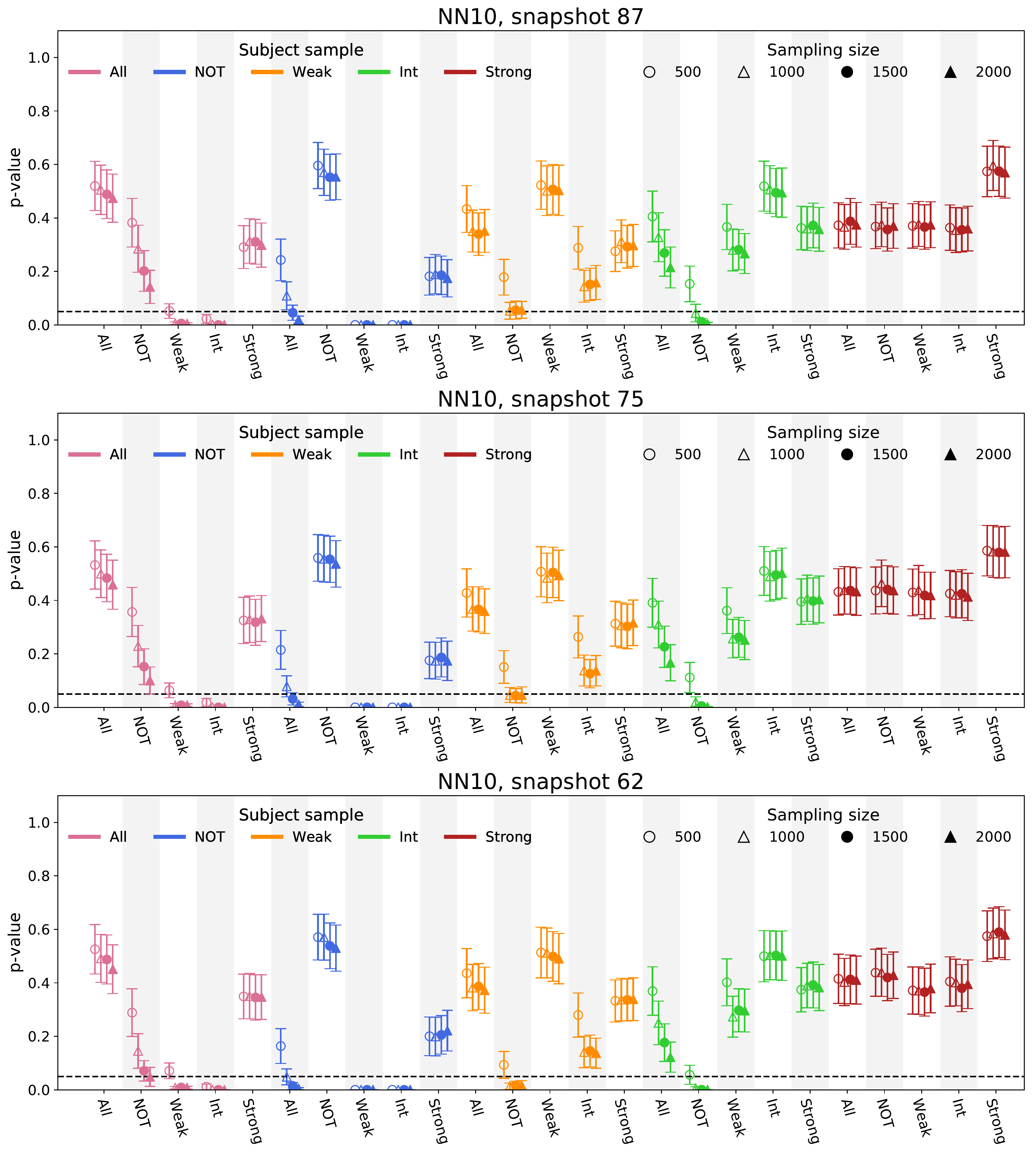}

	\caption{Same as Figure \ref{fig:ks-testsTSLM} but for $D_{10}$ for snapshot 87, 75, and 62}
	\label{fig:ks-testsNN10_p2}
\end{figure*}
\end{appendix}

\bibliography{bibliography}{}
\bibliographystyle{aasjournal}



\end{document}